\documentclass[journal]{IEEEtran}
\usepackage{amsmath,amsfonts}
\usepackage{algorithmic}
\usepackage{algorithm}
\usepackage{array}
\usepackage{textcomp}
\usepackage{url}
\usepackage{verbatim}
\usepackage{graphicx}
\usepackage{cite}
\usepackage{siunitx} 
\usepackage{commath, mathtools} 
\usepackage{bm} 
\usepackage[nolist]{acronym} 
\usepackage{epstopdf} 
\usepackage[hidelinks]{hyperref} 
\usepackage[switch]{lineno} 
\usepackage{xargs} 
\usepackage{multirow} 
\usepackage{lipsum}
\usepackage[table]{xcolor}

\usepackage{pgfplots}
\pgfplotsset{compat=1.9}

\begin{document}
\DeclarePairedDelimiter\floor{\lfloor}{\rfloor}

\def 	\i 	{i} 
\def	\cell	{c} 
\def	\totalcells	{C} 
\def 	\celltarget	{c_\mathsf{target}} 
\def	\Nt	{N_\mathsf{t}} 
\def	\Nu	{N_\mathsf{u}} 
\def 	\fc	{f_\mathsf{c}} 
\def	\slice	{\upsilon} 
\def    \scenario   {\mathsf{sce}}
\def    \marr   {\mathsf{marr}}
\def    \mapf   {\mathsf{mapf}}
\def    \schedslicing   {\mathsf{sched}}
\def    \intentaware    {\mathsf{ia}}
\def    \totalslicetypes  {\Upsilon_{\mathsf{type}}}
\newcommandx{\totalslices}[2][1=,2=]{\Upsilon_{#1}^{#2}} 
\def    \activemet {\mathsf{act}}
\def    \activemetunf {\mathsf{actu}}
\def    \highpriority   {\mathsf{hp}}
\def    \highpriorityunf   {\mathsf{hpu}}
\def	\user 	{u} 
\def    \slicegroup     {\mathsf{gr}}
\def    \snr    {\gamma} 
\def    \noisepower {\sigma}
\def    \pathgainshadow {\alpha} 
\def    \effectivechannel   {h} 
\newcommandx{\totalues}[2][1=,2=]{U_{#1}^{#2}} 
\def 	\bandwidth 	{B} 
\def 	\rbgsavailable 	{R} 
\def 	\rbsavailable 	{G} 
\def    \rbs    {g} 
\newcommandx{\rbgvec}[3][1=,2=,3=(\step)]{\bm{R}_{#1}^{#2}{#3}}
\newcommandx{\rbgsallocated}[2][1=,2=]{R_{#1}^{#2}(\step)} 
\def 	\comb 	{\mathsf{comb}}
\def 	\tti 	{t} 
\newcommandx{\timen}[1][1=]{t_{#1}} 
\def 	\step 	{n} 
\newcommandx{\se}[4][1=,2=,3=,4=(\step)]{\mathsf{SE}_{{#1}{#2}}^{#3}#4} 
\newcommandx{\checkviolations}[4][1=,2=,3=,4=(\step)]{\mathsf{cv}_{{#1}{#2}}^{#3}#4} 
\newcommandx{\vecse}[4][1=,2=,3=,4=(\step)]{\bm{\mathsf{SE}}_{{#1}{#2}}^{#3}#4} 
\newcommandx{\buffer}[3][1=,2=,3=\step]{b_{#1}^{#2}(#3)} 
\newcommandx{\bufferocc}[2][1=,2=]{b_{#1}^{\mathsf{occ}}(\step)} 
\newcommandx{\vecbufferocc}[2][1=,2=]{\bm{b}_{#1}^{\mathsf{occ}}(\step)} 
\newcommandx{\totalbuffer}[1][1=,]{b_{#1}^{\mathsf{max}}} 
\newcommandx{\totallat}[1][1=]{l_{#1}^{\mathsf{max}}} 
\newcommandx{\pktsize}[2][1=,2=]{\mathsf{PS}_{#1}^{#2}} 
\newcommandx{\droppedpkts}[3][1=,2=,3=\step]{d_{#1}^{#2}(#3)} 
\def 	\windowssize 	{w} 
\newcommandx{\rcvthr}[3][1=,2=,3=(\step)]{\iota_{#1}^{#2}{#3}} 
\newcommandx{\thr}[3][1=,2=,3=(\step)]{e_{#1}^{#2}#3} 
\newcommandx{\thrcap}[3][1=,2=,3=(\step)]{r_{#1}^{#2}#3} 
\newcommandx{\lat}[3][1=,2=,3=(\step)]{\ell_{#1}^{#2}#3} 
\newcommandx{\latvec}[3][1=,2=,3=(\step)]{\bm{l}_{#1}^{#2}#3} 
\newcommandx{\latelem}[2][1=,2=]{l_{#1}^{#2}} 
\newcommandx{\pktloss}[3][1=,2=,3=(\step)]{p_{#1}^{#2}#3} 
\newcommandx{\thrlong}[3][1=,2=,3=(\step)]{g_{#1}^{#2}#3} 
\newcommandx{\thrperc}[3][1=,2=,3=(\step)]{f_{#1}^{#2}#3} 
\newcommandx{\fifthperc}[1]{P_{5\%}#1} 
\newcommandx{\muslice}[1][1=]{\mu_{#1}} 
\newcommandx{\sigmaslice}[1][1=]{\sigma_{#1}^2} 
\newcommandx{\normaldist}[1][1=]{\abs{\mathcal{N}(\muslice[#1], \sigmaslice[#1])}} 
\newcommandx{\intentdrift}[3][1=,2=,3=(\step)]{i_{#1}^{#2}#3} 
\newcommandx{\activemetric}[3][1=,2=,3=(\step)]{m_{#1}^{#2}#3} 
\newcommandx{\overindicator}[3][1=,2=,3=(\step)]{q_{#1}^{#2}#3} 
\def \overrate {\zeta} 
\newcommandx{\slicepriority}[1][1=\slice]{p_{#1}} 
\newcommandx{\stepsinterval}[1][1=]{n_{#1}} 
\def 	\samplingtime 	{T_\mathsf{s}} 
\def 	\episodetime 	{T_{e}} 
\newcommandx{\pathloss}[2][1=,2=]{\mathsf{PL}{#2}_{#1}} 
\def 	\umiloss 	{\mathsf{UMi-LOS}} 
\newcommandx{\distance}[1][1=]{d_{#1}} 
\def 	\twod 	{\mathsf{2d}} 
\newcommandx{\breakpoint}[1][1=]{\mathsf{BP}_{#1}} 
\def 	\scalingfactor 	{\mathsf{SF}} 
\newcommandx{\height}[1][1=]{h_{#1}} 
\def 	\basestation 	{b} 
\def 	\environment 	{\mathsf{env}} 
\def 	\threed 	{\mathsf{3d}} 
\def 	\bu 	{b,u} 
\def 	\uminloss 	{\mathsf{UMi-NLOS}} 
\def 	\shadowfading 	{\mathsf{SF}} 
\newcommandx{\rsrp}[2][1=,2=]{\mathsf{RSRP}_{{#1},{#2}}} 
\newcommandx{\lospathcont}[2][1=,2=]{\mathit{\alpha}_{{#1}{#2}}} 
\def	\cluster 	{a} 
\def 	\totalcluster 	{A} 
\def 	\ray	{z} 
\def 	\totalray 	{Z} 
\newcommandx{\transmissionpower}[2][1=,2=]{p_{{#1},{#2}}} 
\def 	\riceanfactor 	{K_\mathsf{R}} 
\newcommandx{\beampattern}[3][1=,2=,3=]{\mathbf{F}_{{#1}{#2}}^{#3}} 
\newcommandx{\initialphase}[4][1=,2=,3=,4=]{\Phi_{{#1}{#2}}^{{#3}{#4}}} 
\newcommandx{\thetaangular}[4][1=,2=,3=,4=]{\theta_{{#1}{#2}}^{{#3}{#4}}} 
\newcommandx{\phiangular}[4][1=,2=,3=,4=]{\phi_{{#1}{#2}}^{{#3}{#4}}} 
\def 	\los 	{\mathsf{LOS}} 
\def 	\aoa 	{\mathsf{AoA}} 
\def 	\eoa 	{\mathsf{EoA}} 
\def 	\aod 	{\mathsf{AoD}} 
\def 	\eod 	{\mathsf{EoD}} 
\def 	\crosspolarization 	{\mathbf{C}} 
\def 	\polarization 	{xy} 
\newcommandx{\kappaconst}[4][1=,2=,3=,4=]{\kappa_{{#1}{#2}}^{{#3}{#4}}} 
\newcommandx{\interference}[4][1=,2=,3=,4=]{I_{{#1}{#2}}^{{#3}{#4}}} 
\def 	\inter	{\mathsf{inter}} 

\def 	\req 	{\mathsf{req}}
\def 	\embb	{\mathsf{embb}}
\def 	\urllc	{\mathsf{urllc}}
\def 	\be	{\mathsf{be}}

\newcommandx{\obsspace}[2][1=, 2=]{\bm{O}_{#1}^{#2}} 
\newcommandx{\slicereq}[1][1=]{\bm{r}_{#1}} 
\newcommandx{\slicemet}[2][1=,2=]{\bm{s}_{#1}^{#2}} 
\newcommandx{\uemet}[1][1=]{\bm{u}_{#1}} 
\def 	\limited	{\mathsf{lim}}
\def 	\interslice	{\mathsf{inter}}
\def 	\intraslice	{\mathsf{intra}}
\newcommandx{\actionspace}[2][1=, 2=]{\bm{A}_{#1}^{#2}} 
\newcommandx{\state}[2][1=,2=]{s_{#1}^{#2}}
\newcommandx{\action}[2][1=,2=]{a_{#1}^{#2}} 
\newcommandx{\objfunction}[1][1=]{J({#1})} 
\def 	\horizon 	{T}
\def 	\temperature 	{\varrho}
\newcommandx{\entropy}[1][1={(\cdot|\state[\tti])}]{\mathcal{H}{#1}}
\def 	\argmin 	{\mathop{\mathsf{arg\,min}}} 
\def 	\dis 	{d} 
\def    \roundsum  {\chi}
\def 	\actionidx 	{\mathsf{index}} 
\newcommandx{\reward}[3][1=,2=,3={(\step)}]{\mathsf{RW}_{#1}^{#2}{#3}} 
\newcommandx{\policy}[1][1=({\state[\tti]}|{\action[\tti]})]{\pi{#1}}
\newcommandx{\rwdweigth}[2][1=,2=]{w_{#1}^{#2}} 
\def 	\stepep 	{\step_\mathsf{ep}} 
\def 	\totalep 	{\mathit{ep}_\mathsf{max}} 
\def 	\eptrain	{\mathit{ep}_\mathsf{train}} 
\def 	\eptest		{\mathit{ep}_\mathsf{test}} 
\def 	\epval		{\mathit{ep}_\mathsf{val}} 
\def 	\epochs 	{\mathit{ec}} 
\def	\rr 	{\mathsf{rr}} 
\def 	\mt 	{\mathsf{mt}} 
\def 	\pf 	{\mathsf{pf}} 
\begin{acronym}[MPC]

    \acro{AoA}{angle of arrival}

    \acro{AoD}{angle of departure}

    \acro{ASHA}{asynchronous successive halving algorithm}

    \acro{A2C}{advantage actor critic}

    \acro{A3C}{asynchronous advantage actor-critic}

    \acro{BE}{best-effort}

    \acro{B5G}{beyond 5G}

    \acro{CDF}{cumulative distribution function}

    \acro{CQI}{channel quality information}

    \acro{CSI}{channel state information}

    \acro{DDPG}{deep deterministic policy gradient}

    \acro{DL}{downlink}

    \acro{DQN}{deep Q-network}

    \acro{DRL}{deep reinforcement learning}

    \acro{EDF}{earliest deadline first}

    \acro{eMBB}{enhanced mobile broadband}

    \acro{EoA}{elevation angle of arrival}

    \acro{EoD}{elevation angle of departure}

    \acro{EURA}{enhanced utilization resource allocation}

    \acro{GBR}{guaranteed bit rate}

    \acro{IaaS}{infrastructure as a service}

    \acro{IBS}{intent-based system}

    \acro{IETF}{internet engineering task force}

    \acro{KPI}{key performance indicator}
    \acrodefplural{KPI}[KPIs]{key performance indicators}

    \acro{LOS}{line-of-sight}

    \acro{LSTM}{long short-term memory}

    \acro{LTE}{long term evolution}

    \acro{MAC}{medium access control}

    \acro{MCS}{modulation coding scheme}

    \acro{MARL}{multi-agent reinforcement learning}

    \acro{MARR}{multi-agent round-robin}

    \acro{MAPF}{multi-agent proportional fair}

    \acro{Mbps}{megabits per step}

    \acro{MDP}{Markov decision process}

    \acro{MIMO}{multiple-input and multiple-output}

    \acro{MTC}{machine-type communication}

    \acro{mMTC}{massive machine-type communication}

    \acro{MT}{maximum throughput}

    \acro{NFV}{network function virtualization}

    \acro{NLOS}{non-line-of-sight}

    \acro{NR}{new radio}

    \acro{OFDM}{orthogonal frequency division multiplexing}

    \acro{ONF}{Open Networking Foundation}

    \acro{PF}{proportional fair}

    \acro{POMDP}{partially observable Markov decision process}

    \acro{PPO}{proximal policy optimization}

    \acro{QCI}{QoS class identifier}
    \acrodefplural{QCI}[QCIs]{QoS class identifiers}

    \acro{QoS}{quality of service}

    \acro{RAN}{radio access network}

    \acro{RB}{resource block}
    \acrodefplural{RB}[RBs]{resource blocks}

    \acro{RBG}{resource block group}
    \acrodefplural{RBG}[RBGs]{resource block groups}

    \acro{RFC}{request for comments}

    \acro{RL}{reinforcement learning}

    \acro{RR}{round-robin}

    \acro{RRM}{radio resource management}

    \acro{RRS}{radio resource scheduling}

    \acro{RSRP}{reference signal received power}

    \acro{SAC}{soft actor-critic}

    \acro{SDN}{software-defined network}

    \acro{SD-RAN}{software-defined radio access network}

    \acro{SE}{spectral efficiency}

    \acro{SISO}{single-input and single-output}

    \acro{SLA}{service-level agreement}

    \acro{SNR}{signal-to-noise ratio}

    \acro{SSR}{SLA satisfaction rate}

    \acro{TDD}{time division duplex}

    \acro{TPE}{tree-structured Parzen estimator}

    \acro{TTI}{transmission time interval}
    \acrodefplural{TTI}[TTIs]{transmission time intervals}

    \acro{UE}{user equipment}
    \acrodefplural{UE}[UEs]{users' equipment}

    \acro{ULA}{uniform linear array}
    \acrodefplural{ULA}[ULAs]{uniform linear arrays}

    \acro{URLLC}{ultra-reliable low latency communication}

    \acro{VNF}{virtual network function}
    \acrodefplural{VNF}[VNFs]{virtual network functions}

    \acro{VoLTE}{voice over LTE}

    \acro{ZOA}{zenith angle of arrival}

    \acro{ZOD}{zenith angle of departure}
\end{acronym} 
\newcommand{\todo}[1]{\textcolor{red}{TODO: #1}}

\title{Intent-based Radio Scheduler for RAN Slicing:\\Learning to deal with different network scenarios}

\author{Cleverson V. Nahum, Salvatore D’Oro, Pedro
Batista, Cristiano B. Both, Kleber V. Cardoso,\\Aldebaro Klautau, \IEEEmembership{Senior Member, IEEE}, and Tommaso Melodia, \IEEEmembership{Fellow, IEEE}.
\thanks{Cleverson V. Nahum and Aldebaro Klautau are with Federal
University of Par\'a, Bel\'em, PA, Brazil (e-mail: {cleversonahum, aldebaro}@ufpa.br). Pedro Batista is with Ericsson Research, Stockholm, Sweden
(e-mail: pedro.batista@ericsson.com). Salvatore D'Oro and Tommaso Melodia are with the Institute for the Wireless Internet of Things, Northeastern University, Boston, MA, USA. (e-mail: s.doro@northeastern.edu). Kleber V. Cardoso is with Universidade Federal
de Goi\'as, Goi\^ania, GO, Brazil. (e-mail: kleber@inf.ufg.br). Cristiano B. Both is with University of Vale do Rio dos Sinos (UNISINOS), São Leopoldo, Brazil. (e-mail: cbboth@unisinos.br).}}

\markboth{XXXXXXXXX, VOL.~XX, NO.~YY, MONTH
    ZZZZ}%
{Cleverson V. Nahum \MakeLowercase{\textit{et al.}}: Intent-based Radio Scheduler for RAN Slicing: Learning to deal with different network scenarios}

\maketitle

\begin{abstract}
The future mobile network has the complex mission of distributing available radio resources among various applications with different requirements. The \acl{RAN} slicing enables the creation of different logical networks by isolating and using dedicated resources for each group of applications. In this scenario, the \ac{RRS} is responsible for distributing the radio resources available among the slices to fulfill their \ac{SLA} requirements, prioritizing critical slices while minimizing the number of intent violations. Moreover, ensuring that the \ac{RRS} can deal with a high diversity of network scenarios is essential. Several recent papers present advances in machine learning-based RRS. However, the scenarios and slice variety are restricted, which inhibits solid conclusions about the generalization capabilities of the models after deployment in real networks. This paper proposes an intent-based \ac{RRS} using \acl{MARL} in a \ac{RAN} slicing context. The proposed method protects high-priority slices when the available radio resources cannot fulfill all the slices. It uses transfer learning to reduce the number of training steps required. The proposed method and baselines are evaluated in different network scenarios that comprehend combinations of different slice types, channel trajectories, number of active slices and \acp{UE}, and \ac{UE} characteristics. The proposed method outperformed the baselines in protecting slices with higher priority, obtaining an improvement of \(40\%\) and, when considering all the slices, obtaining an improvement of \(20\%\) in relation to the baselines. The results show that by using transfer learning, the required number of training steps could be reduced by a factor of eight \(8\) without hurting performance.
\end{abstract}


\IEEEpeerreviewmaketitle

\section{Introduction}
\acresetall

\IEEEPARstart{T}{\lowercase{he}} {6G} networks will support various applications thanks to new technologies and architectures designed to improve network capacity and provide higher throughput, lower latency, and increased reliability~\cite{jiang2021road}. Some applications that will benefit from 6G are smart healthcare, extended reality, virtual reality, holographic communication, and cloud gaming~\cite{quy2023innovative,jiang2021road,lin2022overview}. Technologies such as network slicing, artificial intelligence, and advanced resource allocation techniques are key enablers and are essential for providing network functionality to meet application requirements while improving resource utilization efficiency~\cite{jiang2021road,quy2023innovative}. The network has the complex task of distributing the available resources among various applications, each with different requirements, while improving resource utilization efficiency and providing guarantees of \ac{SLA} fulfillment.

\begin{figure}[t]
    \centering
    \includegraphics[width=\columnwidth]{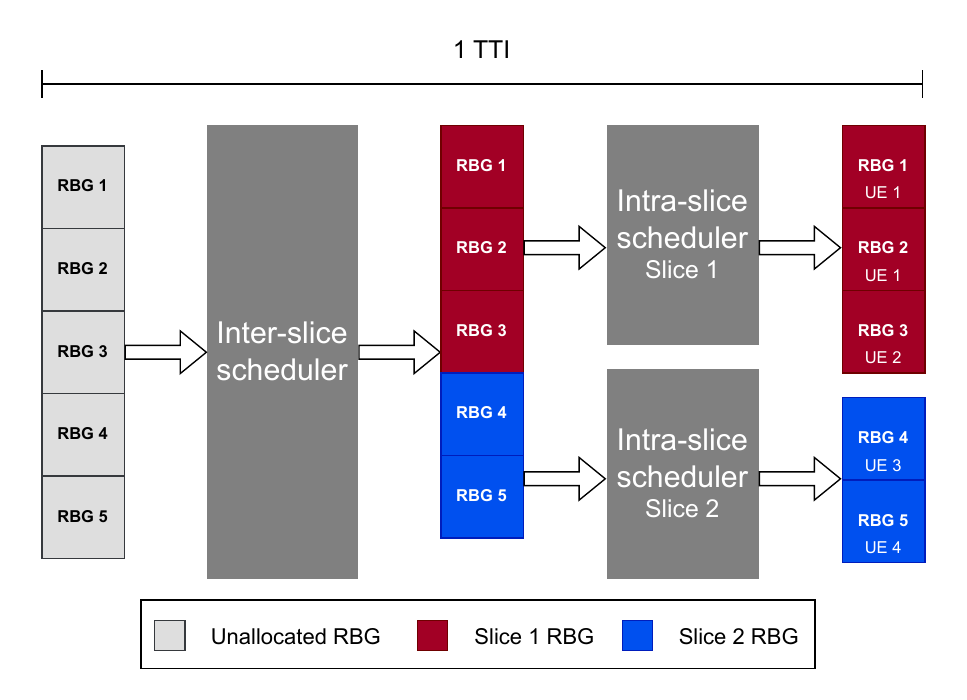}
    \caption{Example of RRS executed over the duration of a \acl{TTI}~(TTI) in a RAN slicing scenario with \(5\) radio resources (RBGs), and \(2\) slices containing \(2\) UEs each. The inter-slice scheduler distributes the available radio resources among slices, and then the intra-slice scheduler distributes the respective radio resources among the \acp{UE} associated with the specific slice.}
    \label{fig:scheduling}
\end{figure}

Network slicing is a technology that provides service customization, isolation, and multi-tenancy support on a shared physical network infrastructure, enabling the logical and physical separation of network resources~\cite{afolabi2018network}. It allows the implementation of independent logical networks adapted for each slice's characteristics and requirements, enabling the specialization of the network to meet the slice's objectives. An end-to-end network slicing involves creating slices in the \ac{RAN}, transport network, and core network domains. One of the \ac{RAN} slicing functions is deploying a \ac{RRS} to allocate radio resources to fulfill the slice requirements defined in the \ac{SLA}. The \ac{RRS} task in a scenario with \ac{RAN} slicing can be split between inter- and intra-slice schedulers~\cite{khodapanah2019radio} as illustrated in Fig.~\ref{fig:scheduling}. The inter-slice scheduler allocates dedicated radio resources for the different slices. To conclude the task, the intra-slice scheduler allocates, in each slice, the resources assigned by the inter-slice scheduler to the \acp{UE}.

A modern \ac{RRS} deals with the whole diversity of applications that the 5G and \ac{B5G} networks can accommodate, being responsible for distributing the available radio resources among the slices to fulfill their requirements and prioritizing critical slices while minimizing the number of \ac{SLA} violations. The \ac{RRS} needs to be aware of the diverse requirements of the different 6G applications. The slice consumer declares the communication service requirements to the operator in an \ac{SLA} through network performance attributes, such as throughput, latency, and reliability requirements~\cite{gsma-ns_req}. Another alternative is to declare these communication service requirements in network slice intents in an intent-based system.
An intent-based system handles intents via a closed-loop process where intents formally specify requirements, goals, and constraints given to a technical system~\cite{tmforum2022}. Using the \ac{RRS} as an intent-based system, slice intents (requirements and goals) can be provided via a common intent model~\cite{tmforum2022A}. In this context, the intent-based \ac{RRS} allocates radio resources to fulfill the received intents. Instead of defining the \ac{RRS} policy to deal with a specific group of slices, intent-based \ac{RRS} receives the intents and implements a policy to fulfill the requested intents. It is possible to add more slices and applications to the system by specifying or updating intents, and the intent-based \ac{RRS} is expected to automatically adapt its policy to meet the new slice's intents.

When considering \ac{RRS} methods for \ac{RAN} slicing scenarios, data-driven approaches have gained increasing attention due to their ability to directly build knowledge about the network from data without the need for statistical models of the system~\cite{wang2019machine}. Machine learning techniques, particularly \ac{RL}, can learn from network data and create flexible policies to deal with the wide variety of \ac{RRS} scenarios in \ac{B5G} networks~\cite{calabrese2018learning,lin2020artificial}. There are different \ac{RRS} methods using \ac{RL} for \ac{RAN} slicing~\cite{nahum2023intent,polese2021colo,yan2019intelligent,mei2021intelligent, li2020lstm,abiko2020flexible,alcaraz2022model,hua2019gan,raftopoulos2024drl}. The methods presented in \cite{polese2021colo,yan2019intelligent,mei2021intelligent} focus on maximizing/minimizing specific network metrics, such as maximizing the slice throughput or minimizing the transmission delay. These works, however, do not consider the case of minimum performance guarantees and, therefore, are incompatible with an intent-based system as they do not target fulfilling specific network requirements. The works~\cite{li2020lstm, abiko2020flexible,alcaraz2022model, hua2019gan,raftopoulos2024drl} are closer to our proposed method in the sense that they support an \acl{SSR} approach in which the network slice objectives are specified. However, they do not provide intent prioritization mechanisms or track intent drift to avoid future intent violations.  Our prior work~\cite{nahum2023intent} was the first to propose intent-aware \ac{RRS} designed to work as an intent-based system for \ac{eMBB}, \ac{URLLC}, and \ac{BE} slices. However, \cite{nahum2023intent} utilizes predefined weights to prioritize intents and always considers the same fixed group of slice types (\ac{eMBB}, \ac{URLLC}, and \ac{BE}). This is a key limitation that is also observed in previous \ac{RRS} works.

In summary, the works in \cite{nahum2023intent,polese2021colo,yan2019intelligent,mei2021intelligent, li2020lstm,abiko2020flexible,alcaraz2022model,hua2019gan,raftopoulos2024drl} do not address the issue of how the \ac{RL} models generalize to diverse and time-varying network scenarios. Specifically, they assess performance using simulations that define a fixed group of slice types, usually \ac{eMBB}, \ac{URLLC}, and \ac{mMTC}, and design \acp{RRS} that are trained to handle specific network conditions (e.g., channel conditions, network load, traffic profiles). In essence, the state-of-art methods were not evaluated when dealing with previously unseen and different network scenarios. Therefore, the previous literature does not provide clear guidelines on how to tackle scenarios that go beyond those considered in relatively restricted simulations.

However, when the goal is to deploy the novel \ac{RL}-based \ac{RRS} method for \ac{RAN} slicing into a real \ac{B5G} network, it is essential to assess the method's capacity to generalize or be utilized for different network scenarios. The \ac{RRS} performance needs to be systematically under conditions such as a varying number of active slices, \acp{UE}, and diverse channel characteristics. Adopting a more adequate assessment methodology, in this work, we focus on developing a single \ac{RRS} method that is able to perform well across different network scenarios to provide a solution for production cellular networks.

We propose an intent-based \ac{RRS} using \ac{MARL} to perform inter- and intra-slice scheduling in a \ac{RAN} slicing scenario. We consider network scenarios that comprehend combinations of different slice types, channel conditions, number of active slices, and \acp{UE}, and \ac{UE} characteristics. Our method utilizes an \ac{RL} agent for inter-slice scheduling, distributing radio resources among the slices. It uses a \ac{MARL} with shared parameters to the intra-slice schedulers, where each slice has its intra-slice scheduler with an \ac{RL} agent, selecting a scheduling algorithm among round-robin, proportional-fair, and maximum throughput to distribute the assigned radio resources among the slice \acp{UE}. 
First, the intent-based \ac{RRS} learns to fulfill the slice intents of the high-priority slices and then learns to meet other regular slices. Prioritizing high-priority slices is implemented in the reward mechanism without needing weight optimization for each network scenario in opposition to~\cite{nahum2023intent}. 

To assess the effectiveness of the proposed method under varying conditions and network configurations, we developed a realistic simulator that relies on the QuaDRiGa channel simulator to generate spatially consistent channels, various traffic models, and \acp{UE} characteristics according to the types of slices. The proposed method and baselines are tested in different network scenarios to assess their capacity to deal with various applications and intents. We formulate three different evaluation scenarios: (i) Training several agents, each one specialized in handling a specific network scenario; (ii) Training a single agent on all network scenarios; and (iii) Using transfer learning when dealing with unseen network scenarios. The first approach evaluates whether specialized \acp{RRS} can handle unseen channel conditions. The second approach determines the generalization capabilities of \acp{RRS} trained on a large set of conditions. Finally, the third approach aims at understanding if the \ac{RRS} can use experiences gathered from different network scenarios to learn how to handle unseen conditions.

The main article contributions are summarized as follows:
\begin{itemize}
    \item Design and development of an intent-based \ac{RRS} using an \ac{RL} agent handling inter-slice scheduling, and \ac{MARL} with shared parameters to implement intra-slice schedulers. The proposed method prioritizes high-priority slices without optimizing predefined weights for each network scenario.
    \item The proposed method handles various network scenarios, fulfilling their intents and prioritizing the high-priority slices when needed.
    \item Improve the intent-drift reward method proposed in~\cite{nahum2023intent} to observe intent drift variations when intents are fulfilled and avoid future intent violations.
    \item Explore the generalization for diverse network scenarios and the use of previous network scenario experiences in training for unseen network scenarios.
    \item Evaluate the proposed method against baselines using various network scenarios with multiple slice types, number of active slices and \acp{UE}, \acp{UE} characteristics, and channel trajectories.
    \item The simulation code is publicly available to facilitate reproducibility and comparison with other methods.\footnote{\url{https://github.com/lasseufpa/intent_radio_sched_multi_slice}}
\end{itemize}

This article is organized as follows: Section~\ref{sec:related_works} extends the brief discussion about the literature already presented. It describes the related work, emphasizing the \ac{RL} \ac{RRS} methods for \ac{RAN} slicing. It also compares the main contributions of this paper concerning previous work. Section~\ref{sec:system_model} presents the adopted communication system model and \ac{RRS} system in a scenario with \ac{RAN} slicing. Section~\ref{sec:proposed_method} presents the proposed intent-based \ac{RRS} using a \ac{MARL} agent to perform inter- and intra-slice allocation. Section~\ref{sec:results} presents the results obtained in three different evaluation scenarios that assess the capacity of the proposed method to deal with varying network scenarios. Finally, Section~\ref{sec:conclusion} summarizes the main results and discusses the challenges of future work.

\section{Related work}
\label{sec:related_works}

The ability of data-driven approaches to learn models directly from the data produced by the network is extremely valuable when considering \ac{RAN} functions. The \ac{RAN} is a data-rich environment that collects data through radio measurements as well as user devices and core network~\cite{wang2019machine}. Data-driven \ac{RRS} methods usually utilize \ac{RL} techniques that can learn from large amounts of data efficiently and that do not require correct pre-computed labels~\cite{calabrese2018learning}.

The employment of \ac{RRS} using \ac{RL} techniques for \ac{RAN} slicing was approached in related works~\cite{nahum2023intent,polese2021colo,yan2019intelligent,mei2021intelligent, li2020lstm,abiko2020flexible,alcaraz2022model,hua2019gan,raftopoulos2024drl}. The solutions proposed in~\cite{polese2021colo,yan2019intelligent,mei2021intelligent,li2020lstm} focus on maximizing/minimizing network metrics, such as maximizing the achieved throughput of \ac{eMBB} slices and minimizing buffer occupancy or latency of \ac{URLLC} slices. For example, work~\cite{polese2021colo} proposes an O-RAN compliant \ac{RL} \ac{PPO} method for both inter- and intra-slice schedulers. Considering the types of slices of \ac{eMBB}, \ac{URLLC}, and \ac{mMTC}. The inter-slice scheduler using \ac{RL} is responsible for selecting the number of \acp{RBG} for each slice while it also selects the intra-slice scheduling algorithm for each slice among the options round-robin, proportional-fair, and maximum-throughput. The implemented reward function, responsible for guiding the \ac{RL} agent learning process, focuses on maximizing the throughput rate to the \ac{eMBB} slice and the transport blocks to the \ac{mMTC} slice while minimizing the buffer size to the \ac{URLLC} slice.

The problem with \ac{RRS} methods focusing on the minimization/maximization of network metrics is the unclear slice objectives. There is no specification of slice intents through network requirements that enables verifying whether the intents have been fulfilled. For example, when the \ac{RRS} method in~\cite{polese2021colo} maximizes the throughput rate to the \ac{eMBB} slice, it is impossible to verify if the technique is reaching its maximization objective because there is no specification of a minimum throughput rate to serve the \ac{eMBB} slices. Therefore, the throughput rate value obtained by the method varies according to the network conditions. It is difficult to verify whether the maximization objective is met since there is no optimum method for comparison. Moreover, when considering intent-based systems based on TM Forum~\cite{tmforum2022}, these minimization/maximization objectives are incompatible with the intent's definition of clearly specifying the requirements the intent-based system should fulfill.

The prior work~\cite{alcaraz2022model,hua2019gan,raftopoulos2024drl,abiko2020flexible} consider \ac{RL} \acp{RRS} with reward functions based on the \acl{SSR}. These methods specify the slice objectives using \ac{QoS} metrics in a \ac{SLA}. The work~\cite{hua2019gan} considers the maximization of spectral efficiency while meeting throughput and latency requirements in a scenario with \ac{VoLTE}, Video, and \ac{URLLC} slices. In~\cite{raftopoulos2024drl}, the presented method considers the latency requirements for two slices with different latency requirement values specified in the \ac{SLA} while minimizing the number of allocated \acp{RBG}, reducing energy consumption. In~\cite{alcaraz2022model}, it considers the \ac{eMBB} and \ac{mMTC} slices. The \ac{eMBB} requirements are the maximum average queue per \ac{GBR} and non-\ac{GBR}, the authorized and compliant capacity for \ac{GBR} in \acp{RB} per subframe. The \ac{mMTC} slice considers the maximum delay per user. The results are assessed in four different network scenarios that comprehend a maximum of five slices with different combinations of the same \ac{eMBB} and \ac{mMTC} slice types. In~\cite{abiko2020flexible}, it defines the throughput and latency requirements for each slice in a network scenario with five slices representing five different types of applications: messaging, app, audio, video, and best effort.

Although the related works~\cite{alcaraz2022model,hua2019gan,raftopoulos2024drl,abiko2020flexible} provide mechanisms for dealing with the slice requirements defined in an \ac{SLA}, they are still insufficient to provide support for slice intents when considering an intent-based system. In~\cite{hua2019gan,alcaraz2022model,abiko2020flexible}, the \ac{SLA} violations are represented as a binary value indicating the fulfillment or not of a requested \ac{QoS} requirement. However, there is no indication of how far the \ac{RRS} agent is from fulfilling their requirements, which is an essential point of intent-based systems~\cite{tmforum2022,rfc9315}, usually represented by an intent drift~\cite{nahum2023intent}, since it gives the \ac{RRS} agent the possibility to assess whether a given action improves/deteriorates the current slice condition. In addition, it enables the report of a more accurate status to the intent owners interested in the slice intent fulfillment. Another critical aspect left out is the slice intent prioritization to cope with high-priority slices in scenarios where the experienced channel capacity is insufficient to fulfill all the requested slice intents.

Our previous work~\cite{nahum2023intent} proposes the first intent-aware \ac{RRS} using \ac{RL} for \ac{RAN} slicing with support for \ac{eMBB}, \ac{URLLC}, and \ac{BE} slice types. It focuses on fulfilling the slice intents defined in a common intent model~\cite{tmforum2022A} and allows for changing the slice intents without retraining the \ac{RRS} agent. The proposed intent-drift reward offers the \ac{RL} agent a distance to fulfill the intents, which allows learning to minimize the distance even when the available radio resources are insufficient to fulfill all the requested intents. It also provides a weight-based prioritization for slice intents to protect high-priority intents in scenarios with high concurrency for radio resources.

Despite the support introduced for slice intents in~\cite{nahum2023intent}, it does not investigate the generalization capabilities of the proposed method to different network scenarios. The \acp{UE} characteristics and mobility pattern do not change in the episodes, and the trained \ac{RL} policy can only deal with variations in the slice intents of the same slice types predefined in the simulation. In addition, using weights to define intent priorities requires joint training and optimization to find the best weight combinations for each network scenario. Therefore, whenever the method is implemented in a different network scenario, the predefined weights must change to reflect the new intent priorities. The proposed intent-drift reward accounts only for the distance to fulfill requirements when slice intents are not fulfilled. However, there is no indication of degrading performance in fulfilled slice intents (helping to prevent future slice violations). Finally, the presented solution uses a fixed round-robin algorithm in the intra-slice scheduler instead of exploring different strategies that could enhance the \ac{RRS} performance.

The main problems related to the support of slice intents in
~\cite{nahum2023intent,polese2021colo,yan2019intelligent,mei2021intelligent, li2020lstm,abiko2020flexible,alcaraz2022model,hua2019gan,raftopoulos2024drl} are summarized below:
\begin{itemize}
    \item The methods focusing on maximizing/minimizing specific network metrics~\cite{polese2021colo,yan2019intelligent,mei2021intelligent,li2020lstm} do not provide sufficient mechanisms to deal with slice intents since it is impossible to define when a given slice intent is fulfilled. Moreover, these maximization/minimization objectives do not comply with the TM forum definition of intents~\cite{tmforum2022}.
    \item Although related works~\cite{alcaraz2022model,hua2019gan,raftopoulos2024drl,abiko2020flexible} provide \ac{RRS} methods based on the \acl{SSR} with \ac{QoS} requirements, these mechanisms offer incomplete support to slice intents since they do not provide a metric to observe intent performance and visualize improvement/degradation over time. In addition, they also do not provide prioritization mechanisms to protect high-priority slice intents when the amount of radio resources at a given moment of the network conditions is insufficient to fulfill all the slice intents.
    \item Our previous work~\cite{nahum2023intent} defines an intent-aware \ac{RRS} using \ac{RL} for \ac{RAN} slicing. Despite its support for slice intents through the intent-drift reward and weight-based priorities, it lacks support for different network scenarios since it requires redefining the weight values for each intent in each network scenario, and there is no clear direction on how to adapt these methods for different combinations of slice types other than the predefined ones.
    \item None of the related works~\cite{nahum2023intent,polese2021colo,yan2019intelligent,mei2021intelligent, li2020lstm,abiko2020flexible,alcaraz2022model,hua2019gan,raftopoulos2024drl} discuss the usage of the proposed methods in different network scenarios in which the slice types, number of active slices, number of \acp{UE} in the slices, \acp{UE} properties, and channel characteristics vary. There is no investigation about the generalizability of the presented methods to different network scenarios or even the use of previously learned experiences in network scenarios that were not seen during the training.
\end{itemize}

Concerning the highlighted issues above, we propose an intent-based \ac{RRS} for \ac{RAN} slicing utilizing \ac{MARL}. We improve the calculation of intent drift reward from~\cite{nahum2023intent} to refine violation prevention and design a \ac{RRS} with homogeneous entries and outputs to support different network scenarios. We investigate the generalizability of the proposed method and baselines in network scenarios not seen in the training. We also explore the transfer learning from different network scenario experiences to fine-tune the agent to deal with unseen network scenarios.

\section{Communication system model and problem formulation}
\label{sec:system_model}

To simplify notation, in the following, we consider a \ac{SISO} system with a single base station operating at carrier frequency \(\fc\) and providing service to \(\slice = 1,2,3,\dots,\totalslices\) \ac{RAN} slices. The base station has \(\user=1,2,3,\dots,\totalues\) \acp{UE} connected at the same time, where each \ac{UE} \(\user\) has a single antenna and is assigned to a specific slice \(\slice\). Each slice \(\slice\) contains a set of \(\totalues[\slice]\) \acp{UE}.  
Despite the \ac{SISO} assumption, we would like to mention that our model is general and applies to other RF transmission schemes, such as \ac{MIMO}, without changes in our proposed system. 

The base station has a \(\bandwidth\)~\si{\mega\hertz} bandwidth, divided into \(\rbsavailable\) \acp{RB}, and \acp{RB} are grouped into \(\rbgsavailable\) \acp{RBG} that is considered the minimum radio resource allocation unit in the scheduling system. The \ac{TTI}~\(\tti\) is measured in~\si{\milli\second} and represents the minimum time unit considered in the scheduling process. In each \ac{TTI}, the inter- and intra-slice schedulers allocate all \acp{RBG} in a specific simulation step \(\step\). Each step \(\step\) takes \(\tti\)~\si{\milli\second} with \(\timen[\step]=\tti\cdot\step\) representing the time from step \(1\) to \(\step\). The \ac{RRS} system with \ac{RAN} slicing distributes the available \(\rbgsavailable\) \acp{RBG} to the active slices \(\totalslices[\activemet]\). We assume an uplink \ac{RRS} formulation where the base station defines the \acp{RBG} each \ac{UE} uses, but the proposed method also applies to downlink.

\subsection{Channel modeling}

We use a \ac{TDD} transmission protocol that receives pilots from \acp{UE} sent through the uplink to obtain the base station's \ac{CSI}. Each base station obtains a perfect channel estimation for each \ac{UE}. \acp{UE} have their spectral efficiency values varying with time and frequency, which is different from \cite{nahum2023intent,chen2021adaptive,polese2021colo,feng2020dynamic}, where the same \ac{UE}'s spectral efficiency value is adopted for all \acp{RBG}.

We use QuaDRiGa software~\cite{jaeckel2014quadriga} to consistently generate \ac{UE} channels in space and frequency. It generates spatially and correlated channels from statistical models, including experimentally validated channel models. We use the 3GPP 38.901 UMa~\cite{zhu20193gpp} statistical models based on dual-slope path loss with significant inter-parameter correlations. Furthermore, given that the base station allocates the \(\rbs\)-th \ac{RB} to the \(\user\)-th \ac{UE}, the \ac{SNR} perceived by the \ac{UE} can be expressed as follows
\begin{equation}\label{Eq:SNR}
\snr_{\user,\rbs} = \frac{\pathgainshadow_{\user} \ \transmissionpower[\user][\rbs] \  \lvert \effectivechannel_{\user,\rbs} \rvert^{2}}{\noisepower_{\user}^{2}},
\end{equation}
where \(\transmissionpower[\user][\rbs]\) is the allocated transmit power to the \ac{UE} \(\user\) in the \(\rbs\)-th \ac{RB}, \(\pathgainshadow_{\user}\) is the effect of path gain and shadowing experienced by the \(\user\)-th \ac{UE}, \(\effectivechannel_{\user,\rbs}\) is the effective channel for \ac{UE} \(\user\) in the \(\rbs\)-th \ac{RB}, and \(\noisepower_{\user}\) is the noise power experienced by the \(\user\)-th \ac{UE}. This way, the spectral efficiency \(\se[\user,\rbs][][]\) to \ac{RB} \(\rbs\) and \ac{UE} \(\user\) is defined as
 \begin{equation}
\se[\user,\rbs][][]=\log_2(1+\snr_{\user,\rbs}).
\end{equation}
In this work, we employed QuaDRiGa \ac{LOS} and \ac{NLOS} channels, as documented in~\cite{QuadrigaRep}. 

\subsection{Radio resource scheduling with RAN slicing}

The number of \acp{RBG} allocated by the inter-slice scheduler for each slice in a simulation step \(\step\) is represented in the vector
\begin{equation}
    \label{eqn:rbs_inter_vec}
    \rbgvec[][\interslice] = [\rbgsallocated[1], \rbgsallocated[2], \dots, \rbgsallocated[\totalslices]],
\end{equation}
with \(\rbgsallocated[\slice]\) representing the number of \acp{RBG} allocated to slice \(\slice\) at step \(\step\). We assume, for simplicity, that all \acp{RBG} are allocated during the scheduling process. Therefore, the \ac{RRS} obeys to
\begin{equation}
    \sum_{\slice=1}^{\totalslices}\rbgsallocated[\slice]=\rbgsavailable
    .
\end{equation}
In case the slice \(\slice\) does not have sufficient data to use all the allocated \acp{RBG}, the extra \acp{RBG} are reserved for the slice but not used. These allocation decisions can be easily converted to the 3GPP specification to the \ac{RRM} policy ratio~\cite{ts-3gpp-28.541}. 

The inter-slice scheduler defines the number of \acp{RBG} \(\rbgsallocated[\slice]\) to the slice \(\slice\) at step \(\step\), and then the intra-slice scheduler allocates this \(\rbgsallocated[\slice]\) \acp{RBG} available among the \(\totalues[\slice]\) slice \acp{UE} in the vector
\begin{equation}
    \label{eqn:rbs_intra_vec}
    \rbgvec[\slice][\intraslice] = [\rbgsallocated[\slice][1], \rbgsallocated[\slice][2], \dots, \rbgsallocated[\slice][{\totalues[\slice]}]],
\end{equation}
where \(\rbgsallocated[\slice][\user]\) represents the number of \acp{RBG} allocated to the \ac{UE} \(\user\) in the slice \(\slice\) at step \(\step\). The intra-slice scheduler also obeys to
\begin{equation}
    \sum_{\user=1}^{\totalues[\slice]}\rbgsallocated[\slice][\user]=\rbgsallocated[\slice]
    .
\end{equation}

The \ac{RRS} performance is evaluated considering the network metrics defined in the proposed slice intents. We define served throughput as the maximum throughput in \ac{Mbps} that could be achieved by a \ac{UE}~\(\user\) in the slice \(\slice\) taking into account the number of \acp{RBG} \(\rbgsallocated[\slice][\user]\) allocated and their spectral efficiency values \(\se[\user]\)

\begin{equation}
    \thrcap[\slice][\user]=
    \floor*{\frac{\bandwidth\sum_{\rbs \in \rbsavailable^{\user}}{\se[\user,\rbs]}}
        {\pktsize[\slice]\ \rbsavailable\ 10^6}}\pktsize[\slice] ,
\end{equation}
where \(\rbsavailable^{\user}\) represents the \acp{RB} allocated to a \ac{UE} \(\user\), \(\pktsize[\slice]\) is the packet size in \SI{}{bits} for \acp{UE} in the slice \(\slice\). We do not consider the \ac{MCS} from each \ac{RB} in the throughput calculation for clearness.

The data available to send in the \ac{UE} buffer limits the served throughput. Therefore, the effective throughput \(\thr[\slice][\user]\) represents the data throughput sent over the network
\begin{equation}
    \thr[\slice][\user]=\min(\thrcap[\slice][\user], \buffer[\user]),
\end{equation}
with \(\buffer[\user]\) representing the data available of UE~\(\user\) at step \(\step\) in~\si{\mega\bit}. As a consequence, the effective throughput is always
\(\thr[\slice][\user]\leq\thrcap[\slice][\user]\) with \(\thr[\slice][\user]=\thrcap[\slice][\user]\) when
\(\buffer[\user]\geq\thrcap[\slice][\user]\).

The buffer occupancy rate \(\bufferocc[{\user,\slice}]\) is defined as
\begin{equation}
    \bufferocc[{\user,\slice}]=\frac{\buffer[\user]}{\totalbuffer[\slice]} ,
\end{equation}
where \(\totalbuffer[\slice]\) is the maximum \ac{UE} buffer capacity in slice \(\slice\). Packets are discarded whenever the buffer is full or the packet latency exceeds the maximum allowed latency \(\totallat[\slice]\). For this reason, these packets are accounted for in the dropped data \(\droppedpkts[\user]\) that represents the size of the packets dropped in step \(\step\). The packet loss rate \(\pktloss[\user]\) is calculated over a window interval of \(\windowssize\) steps
\begin{equation}
    \pktloss[\slice][\user]=
    \begin{cases}
        \frac{\sum_{\i=(\step-\windowssize)}^\step
            \droppedpkts[\user][][\i]}{\buffer[\user][][\step-\windowssize]+
        \sum_{\i=(\step-\windowssize)}^\step \rcvthr[\slice][\user][(\i)]}, &
        \text{if } \step\geq\windowssize                              \\
        \frac{\sum_{\i=1}^\step
            \droppedpkts[\user][][\i]}{\buffer[\user][][1]+\sum_{\i=1}^\step
        \rcvthr[\slice][\user][(\i)]},                                      &
        \text{otherwise}
    \end{cases}
    ,
\end{equation}
where \(\rcvthr[\slice][\user]\) is the requested throughput by \ac{UE} \(\user\) in slice \(\slice\) at step \(\step\). The requested throughput depends on the slice \(\slice\) that the \ac{UE} is associated with since the traffic behavior of \acp{UE} of the same slice is similar.

The average time that packets have waited in the buffer of \ac{UE} \(\user\) is represented as
\begin{equation}
    \lat[\user]=\frac{\sum_{\i=0}^{\totallat}
        \i\latvec[\step][\user][(\i)]}{\sum_{\i=0}^{\totallat}
        \latvec[\step][\user][(\i)]} ,
\end{equation}
where \(\latvec[\step][\user][]=[\latelem[0], \latelem[1], \hdots,
\latelem[{\totallat[\slice]}]]\) is a vector with size
\(\totallat[\slice]+1\) representing the packets' latency on buffer for user \(\user\) at step \(\step\), and \(\latelem[{\totallat[\slice]}]\) represents the number of packets that have waited for \(\totallat[\slice]\) \acp{TTI} in the buffer.

Slice metrics average the \acp{UE} metrics associated with the target slice. For example, the effective throughput \(\thr[\slice][]\) for slice \(\slice\) is the average effective throughput from its \acp{UE} \(\totalues[\slice]\).

\begin{figure*}[ht]
    \centering
    \includegraphics[width=\textwidth]{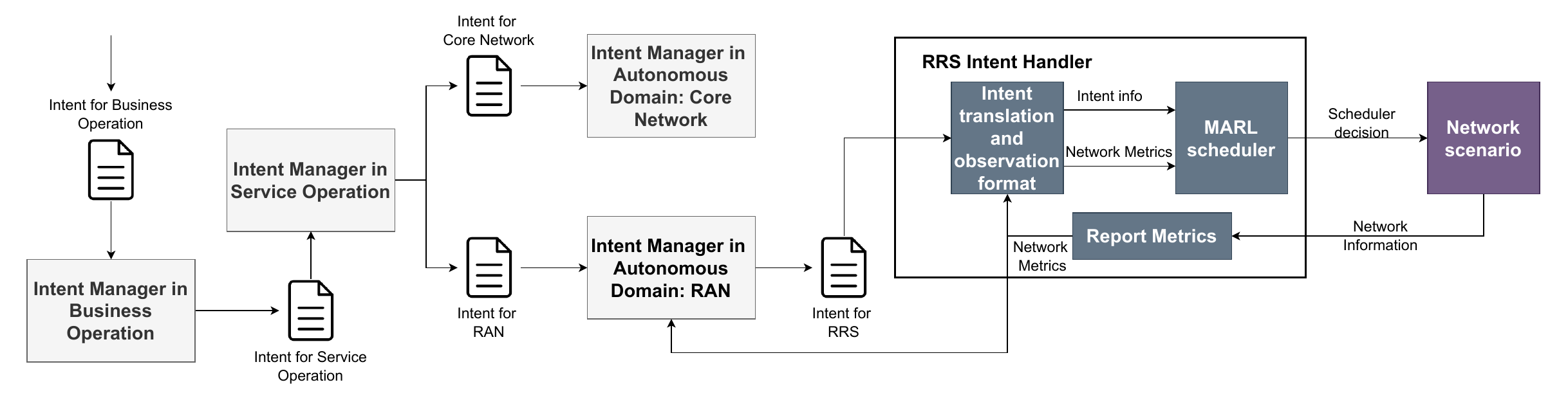}
    \caption{An intent-based system with an \ac{RRS} intent handler responsible for receiving slice intents and network information to generate schedule actions to fulfill the provided intents. The \ac{RRS} intent handler is based on \ac{MARL}.}
    \label{fig:bigpicture}
\end{figure*}

\subsection{Slice intents and requirements}
\label{subsec:intents_requirements}

Related works~\cite{nahum2023intent,polese2021colo,yan2019intelligent,mei2021intelligent, li2020lstm,abiko2020flexible,alcaraz2022model,hua2019gan,raftopoulos2024drl} usually define three different slice types based on 5G use cases: \ac{eMBB}, \ac{URLLC}, and \ac{mMTC}. This definition hides the diversity of the application inside each of these use cases. For example, the \ac{eMBB} use-case can contain applications such as video streaming and cloud gaming, where their network requirements are very distinct. An application running a \(4\)K video streaming on Netflix has a throughput requirement of \SI{15}{Mbps}~\cite{netflix2024internetreq}. At the same time, a Nvidia cloud gaming application has a throughput requirement of \SI{45}{Mbps} and also a latency requirement of \SI{40}{ms} latency for the best experience~\cite{nvidia2024systemreq}. The adopted \ac{RRS} needs to deal with these applications differently to fulfill their intents instead of grouping them in the same slice and fitting them with identical objectives.

We propose a definition of the slice type coupled with end-user applications, and then the \ac{RRS} needs to deal with different network slice intents. The main goal is to handle these slice intents independently of the network's combination of active slice types. To accomplish that, we consider that each type of slice has to define its intents based on three main network metrics: effective throughput, buffer latency, and packet loss rate. Each slice type must have one or more intent definitions, but the intent requirements do not need to consider the same metrics since each application has its characteristics. The diversity of applications is essential to ensure the proposed \ac{RRS} scheme can deal with different types of slices and combinations of active slices.

Each slice \(\slice\) can have up to three different slice intents in each simulation step \(\step\): the effective throughput intent requirement \(\thr[\slice][\req][]\) in \SI{}{Mbps}, the buffer latency \(\lat[\slice][\req][]\) in \SI{}{ms}, and the packet loss rate \(\pktloss[\slice][\req][]\). Each slice \(\slice\) has a binary active intent indicator for effective throughput \(\activemetric[\slice][{\thr[][][]}][]\), buffer latency \(\activemetric[\slice][{\lat[][][]}][]\), and packet loss rate \(\activemetric[\slice][{\pktloss[][][]}][]\), which indicates if the slice considers the target network metric in its intents. Therefore, the slice \(\slice\) intents are fulfilled every time the following conditions are achieved:

\begin{equation}
    \label{eqn:network_reqs}
    \begin{aligned}
        \frac{\sum_{\user=1}^{\totalues[\slice]}{\thr[\slice][\user]}}{\totalues[\slice]} \ge \thr[\slice][\req][], \text{if } \activemetric[\slice][{\thr[][][]}][]=1
        \\
        \frac{\sum_{\user=1}^{\totalues[\slice]}{\lat[\slice][\user]}}{\totalues[\slice]} \le \lat[\slice][\req][], \text{if } \activemetric[\slice][{\lat[][][]}][]=1
        \\
        \frac{\sum_{\user=1}^{\totalues[\slice]}{\pktloss[\slice][\user]}}{\totalues[\slice]} \le \pktloss[\slice][\req][], \text{if } \activemetric[\slice][{\pktloss[][][]}][]=1
    \end{aligned}
\end{equation}

Whenever one or more slice intents are not fulfilled, the slice accounts for a violation. The \ac{RRS} function in an intent-based system with \ac{RAN} slicing is to prevent/minimize the intent violations.

\section{Proposed Intent-based RRS agent using MARL}
\label{sec:proposed_method}

This work proposes an intent-based \ac{RRS} for a scenario with \ac{RAN} slicing using \ac{MARL} based on TM Forum IG1253~\cite{tmforum2022} definitions for an intent-based network. TM Forum defines intent as the formal specification of all expectations, including requirements, goals, and constraints given to a technical system. Furthermore, intents should be expressed with formality and complete semantics and vocabulary, avoiding ambiguities in the meaning of an intent (the sender and receiver of intent must be in perfect agreement about its interpretation). 

Fig.~\ref{fig:bigpicture} shows an intent-based system focusing on the \ac{RRS} for \ac{RAN} slicing. The intent manager in business operations receives a high-level intent that can contain the slice type/application description and generates an intent for service operation containing requirements about service \acp{KPI} and customer experience metrics. The intent manager in the service operation receives the previous intent and creates intents to the \ac{RAN}, transport, and core network intent managers so that the generated intents could satisfy the intent for the service operation. Finally, in the \ac{RAN} domain, the intent manager receives the \ac{RAN} intents and coordinates their different functions, such as the \ac{RRS}, to fulfill the intent for RAN requirements. Communication between intent managers is possible through a common intent model~\cite{tmforum2022A} that specifies the intent description and reports using common vocabulary and semantics.

We propose an \ac{RRS} intent handler that receives the intent for \ac{RRS} (following the common intent model~\cite{tmforum2022A}) specifying the intents for each slice to be fulfilled by the \ac{RRS} operations. The RRS intent handler is responsible for receiving the intent for \ac{RRS} and generating radio scheduler decisions to fulfill the intents or minimize the number of violations in scenarios where the radio resources available are insufficient to meet all the slice's intents. The intent translation and observation format function is responsible for translating the received slice intents to the intent information represented in the effective throughput intent requirement \(\thr[\slice][\req][]\), the buffer latency \(\lat[\slice][\req][]\), and the packet loss rate \(\pktloss[\slice][\req][]\) described in Subsection~\ref{subsec:intents_requirements}. In addition, it also creates a vector containing network metrics and intent fulfillment information to the \ac{MARL} scheduler.

The proposed \ac{MARL} scheduler utilizes the intent information and network metrics to generate a scheduler decision indicating the \acp{RBG} to allocate for each slice's \ac{UE} aiming at fulfilling the slice intents. The \ac{RRS} intent handler applies the scheduler decision in the network scenario, providing updated network information due to the scheduler action. Finally, the report metrics function is responsible for calculating network metrics to report to the \ac{RAN} intent manager, provide information about the slice's intent fulfillment, and also provide network metrics to the observation format used as input in the \ac{MARL} scheduler.

\subsection{Intent-based RRS using MARL}
\label{subsec:intent_based_rrs}

We propose a \ac{MARL} method to perform inter- and intra-slice scheduling in a \ac{RAN} slicing scenario similar to~\cite{nahum2023intent}. The inter-slice scheduler provides an action representing the number of \acp{RBG} for each slice, and the intra-slice scheduler selects an algorithm method among round-robin, maximum-throughput, and proportional-fair~\cite{capozzi2012downlink}. The proposed \ac{MARL} agent aims to fulfill the intents of active slices in the network, considering different scenario combinations, including variations in the number of slices, type of slices, and number of \acp{UE} associated with each slice.

\begin{figure}[t]
    \centering
    \includegraphics[width=\columnwidth]{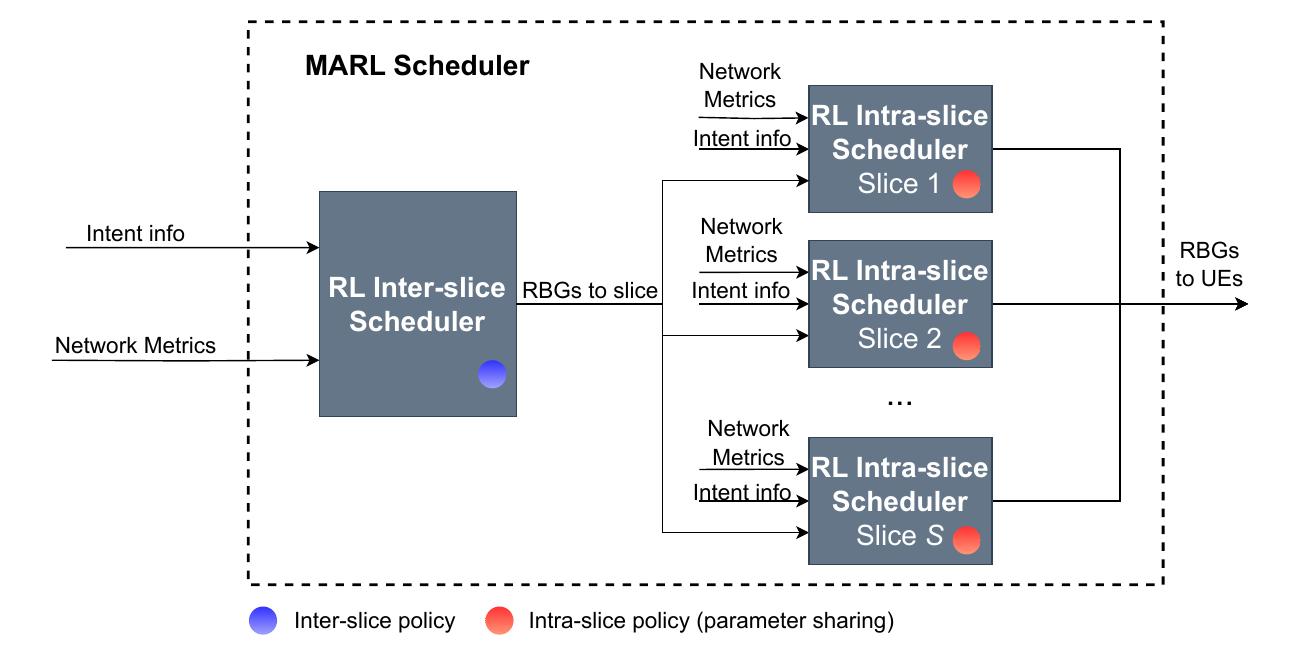}
    \caption{Proposed intent-based \ac{MARL} architecture to perform inter- and intra-slice scheduling in different network scenarios. The \ac{RL} inter-slice scheduler has a dedicated policy, while the \ac{RL} intra-slice schedulers utilize a shared policy.}
    \label{fig:marl_arch}
\end{figure}

We adopt a \ac{PPO} \ac{RL} method for being a well-established method for dealing with continuous control tasks and the stability and reliability of trust-region methods with a less complex implementation~\cite{schulman2017proximal}. The proposed method for \ac{RRS} implements an inter-slice scheduler utilizing a \ac{PPO} \ac{RL} method with a dedicated policy. In contrast, intra-slice schedulers utilize a \ac{MARL} \ac{PPO} with parameter sharing (shared policy)~\cite{terry2020revisiting} as depicted in Fig.~\ref{fig:marl_arch}. Concerning the \ac{RL} inter-slice scheduler, we formulate the system as a \ac{MDP} using a tuple \((S, \actionspace[][\interslice], \reward[][\interslice][], P, \rho_0)\), where \(S\) is the set of all valid states, \(\actionspace[][\interslice]\) is the set of all valid actions for the inter-slice scheduler, \(\reward[][\interslice][]\) is the reward function, \(P\) is the transition probability function, and \(\rho_0\) is the initial state distribution of the system~\cite{sutton2018reinforcement}. In a time step \(\tti\), the agent in a state \(\state[\tti]\) takes action \(\action[\tti]\) and reaches the next state \(\state[\tti+1]\) receiving the reward \(\reward[][\interslice][]\). Rewards are numerical values given to the agent's actions to represent if the chosen action was effective, and the agent aims to maximize the long-term cumulative reward~\cite{sutton2018reinforcement}. In the inter-slice case, the reward represents the fulfillment of the slices' intents. The inter-slice agent follows a policy \(\policy[{(\cdot|\state[\tti])}]\) defined as a distribution over actions given the current state \(\state[\tti]\).

We adopt the \ac{PPO} implementation using a clipped surrogate objective
\begin{equation}
    L_{t}^{\mathsf{clip}}(\theta) = \mathbb{\hat E}_t \!\! \left[ \min ( r_t(\theta) \hat A_t, \text{clip}(r_t(\theta), 1 - \epsilon, 1 + \epsilon) \hat A_t ) \right],
\end{equation}
where the probability ratio 
\begin{equation}
    r_t(\theta)=\frac{\policy[]_{\theta}(\action[\tti]|\state[\tti])}{\policy[]_{\theta\mathsf{old}}(\action[\tti]|\state[\tti])}
\end{equation} represents the current policy \(\policy[]_{\theta}\) changes in relation to the old policy \(\policy[]_{\theta\mathsf{old}}\). The estimated advantage function \(A_t\) utilizes generalized advantage estimation~\cite{schulman2017proximal}, measuring how much better or worse a particular action is compared to the agent’s average performance. It clips the \(r_t(\theta)\) value outside the interval \([1 - \epsilon, 1 + \epsilon]\), where \(\epsilon\) is a hyper-parameter. It takes the minimum between the clipped and unclipped objective, so the final objective is a lower bound on the unclipped objective. Clipping the objective between the defined interval improves the stability and reliability when updating the policy values.

Finally, the \ac{PPO} total loss is
\begin{equation}
    L_{\mathsf{total}}(\theta) = \mathbb{E}_t \left[-L_{t}^{\mathsf{clip}}(\theta) + c_{1}L_{t}^{\mathsf{VF}}(\theta) - c_{2}L_{t}^{\mathsf{entropy}}(\theta) \right],
\end{equation}
which includes the value function loss \(L_{t}^\mathsf{VF}(\theta)\) and an entropy loss \(L_{t}^\mathsf{entropy}(\theta)\) to encourage the exploration, and their respective coefficients \(c_1\) and \(c_2\)~\cite{schulman2017proximal}. The total loss represents the overall objective that the training process seeks to minimize.

The method proposed for the intra-slice scheduler utilizes \ac{MARL} in which there is one \ac{RL} agent for each slice \(\slice\) totaling \(\totalslices\) intra-slice agents. We formulate the \ac{MARL} using a \ac{POMDP} defined by a tuple \((\totalslices,S,\{\actionspace[\slice][\intraslice]\},\{\obsspace[\slice][\intraslice]\}, \{\reward[\slice][\intraslice][]\},T,O)\)~\cite{gupta2017cooperative}, where \(\totalslices\) represents the total number of slices in the system, which is equal to the number of intra-slice agents, \(\{\actionspace[\slice][\intraslice]\}\) is a set of actions for each agent from slice \(\slice\), \(\{\obsspace[\slice][\intraslice]\}\) is a set of observations for each agent from slice \(\slice\), \(\{\reward[\slice][\intraslice][]\}\) is a set of reward signals for each agent from slice \(\slice\), \(T\) and \(O\) are the joint transition and observation models. The parameter-sharing approach is used since the intra-slice scheduler agents are homogeneous, allowing them to share the parameters of a single policy~\cite{gupta2017cooperative}. This allows the policy to obtain the experiences of all agents simultaneously, but it is still different agents since they receive different observations. The \ac{PPO} clipped surrogate objective when using shared parameters to the intra-slice scheduler is
\begin{equation}
        L_{t}^{\mathsf{clip}}(\theta) = \frac{1}{\totalslices} \sum_{\slice=1}^{\totalslices} \mathbb{\hat E}_t \!\! \left[ \hat A_t^\slice \min ( r_t^\slice(\theta), \mathsf{clip}(r_t^\slice(\theta), 1\! -\! \epsilon, 1\! +\! \epsilon))\right].
\end{equation}

\begin{figure}[t]
    \centering
    \includegraphics[width=\columnwidth]{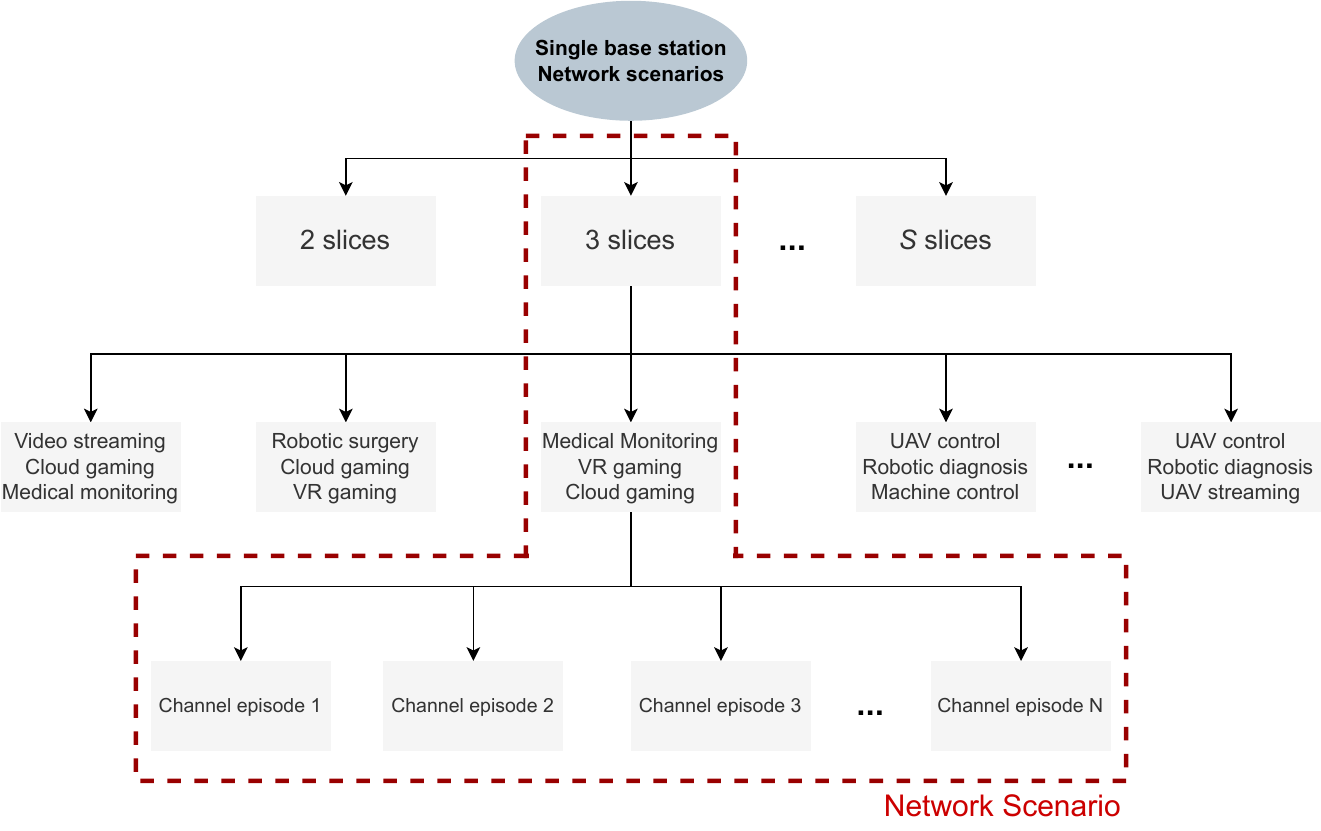}
    \caption{Single base station network scenario possibilities considering a different number of active slices, slice types, and \ac{UE} channel trajectories. A network scenario is a specific combination of slices, slice types, and different channel conditions.}
    \label{fig:network_scenarios}
\end{figure}

We define a network scenario as a specific combination of active slices, slice types, number of \acp{UE} assigned for each slice, and the different \acp{UE} channel trajectories. Fig.~\ref{fig:network_scenarios} illustrates all the possible combinations of our network scenario definition. The variation in the number of \acp{UE} per slice was omitted to simplify the visualization. Related works~\cite{nahum2023intent,polese2021colo,yan2019intelligent,mei2021intelligent, li2020lstm,abiko2020flexible,alcaraz2022model,hua2019gan,raftopoulos2024drl} usually consider a unique network scenario to train and test the designed methods. Considering these training/testing conditions, these methods can deal with any channel episode if we keep the same network scenario characteristics they were intended for. Due to the high diversity of slice types, training and testing the \ac{RRS} methods under different network scenarios is essential to evaluate their ability to deal with various applications and fulfill slice intents.

When using an \ac{RL} method to deal with different network scenarios, the observation space, the action space, and the reward function of the \ac{RL} agent need to be able to deal with this variation since the number of entries in the neural network is fixed~\cite{charu2018neural}. Related works~\cite{nahum2023intent,polese2021colo,yan2019intelligent,mei2021intelligent, li2020lstm,abiko2020flexible,alcaraz2022model,hua2019gan,raftopoulos2024drl} usually consider a set of different variables per slice type, making it impossible to use these methods for different combinations of network scenarios, since a different number of active slices and slice types would lead to a variable number of entries in the \ac{RL} agent, requiring a change in the number of entries in the inputs of the neural network or the reward calculation. 

When based on state-of-art \ac{RL} techniques such as~\cite{haarnoja2018soft,schulman2017proximal}, using the same \ac{RL} \ac{RRS} method for different network scenarios requires a homogeneous input per slice type where each slice type should be represented by the same set of variables, keeping the same position in the entries of the neural network. Therefore, the same neural network structures from \ac{RL} agents could be used for different network scenarios. Moreover, the contribution of the slice to the reward function must be calculated similarly to enable the \ac{RL} agent to understand the different goals needed for each network scenario without changing the \ac{RL} structure. The proposed method obeys these characteristics, enabling its use in various network scenarios.

\subsection{Intent-drift calculation}
An intent drift occurs when a system initially meets the defined intent but gradually, over time, allows its behavior to change or be affected until it no longer does or does so in a less effective manner~\cite{rfc9315}. Related work~\cite{nahum2023intent} represents the intent drift as a value between \(-1\) and \(0\) with \(0\) representing the fulfilled intention and \(-1\) representing the worst performance (most considerable distance from the current metric value and the requirement). These intent drift values show how distant the \ac{RL} agent policy is from fulfilling the requirements. However, it also lacks information on performance degradation when intents are still fulfilled. For example, a slice with an effective throughput intent of \(\thr[\slice][\req][]=10\)~\SI{}{Mbps} might receive \SI{11}{Mbps} in a given moment. However, changes in network conditions can slightly decrease effective throughput, leading to an unfulfilled intent in the future. The representation of intent drift in~\cite{nahum2023intent} does not account for this performance degradation that could give the \ac{RL} agent important information about avoiding future intent violations.

\begin{figure}[t]
    \centering
    \includegraphics[width=\columnwidth]{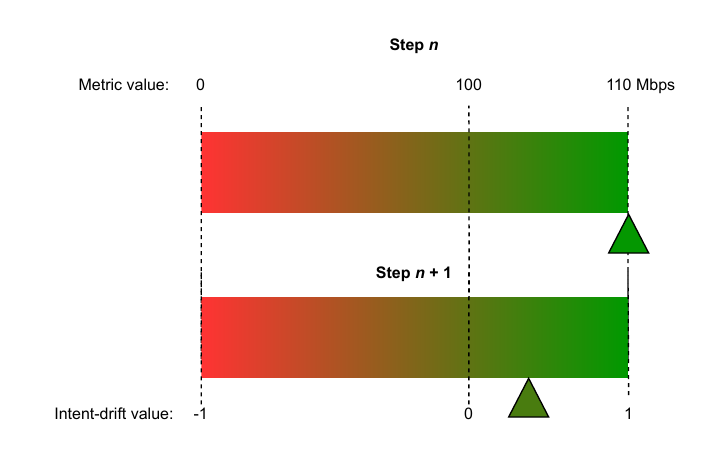}
    \caption{Intent-drift example for an effective throughput intent with a requirement of 100~Mbps and an overfulfillment rate of \(10\)\%. In step \(\step\), the served throughput is equal to or greater than the requested intent requirement. The served throughput decreases in step~\(\step+1\), but the intent is still fulfilled.}
    \label{fig:intent_drift}
\end{figure}

In this work, we represent the intent drift as a distance to fulfill the intent requirements (similarly as~\cite{nahum2023intent}) and as a representation of degrading metrics even if the slice intents are still fulfilled. In addition to the distance to fulfill the intent requirements, we also account for the distance between the requirement and an over-fulfillment state, represented as a percentage above the requirement. Fig.~\ref{fig:intent_drift} shows an example of the intent drift values for an effective throughput intent with a requirement of \SI{100}{Mbps} and an over-fulfillment rate of \(10\%\). Every time the effective throughput is under the specified requirement of \SI{100}{Mbps}, the intent drift accounts for a value between \(-1\) and \(0\). If effective throughput is a value between \SI{100}{Mbps} and \(110\) (over-fulfilled throughput), a value between \(0\) and \(1\) is taken into account. In case the effective throughput is greater than \SI{110}{Mbps}, the intent drift is \(1\). In step \(\step\), the intent drift is \(1\) since it is receiving an effective throughput equal to or greater than \SI{110}{Mbps}. However, in step \(\step+1\), the effective throughput decreased in value but still met the intent requirements of \SI{100}{Mbps}. The intent drift can provide information about performance degradation even in fulfilled intents so the \ac{RL} agent can avoid unsafe fulfillment zones that can lead to unfulfilled intents in the future.

We calculate the intent drift for three intent requirements: effective throughput, buffer latency, and packet loss rate. The intent drift for effective throughput \(\intentdrift[\user][{\thr[][][]}][]\) in a simulation step \(\step\) for \ac{UE} \(\user\) is

\begin{equation}
    \label{eqn:intent_thr_ue}
    \intentdrift[\user][{\thr[][][]}] = 
    \begin{cases}
        \frac{\thr[\user] - \thr[\user][\req][]}
        {\thr[\user][\req][] \overindicator[\user][{\thr[][][]}]}, & \!\!\!\! \text{if } \thr[\user]<\thr[\user][\req][] (1+\overindicator[\user][{\thr[][][]}]) \\ & \text{and} \  \bufferocc[\user] > 0 \\
        1,\,                      & \!\!\!\! \text{Otherwise}
    \end{cases}
    ,
\end{equation}
where the over-fulfilled requirement indicator \(\overindicator[\user][{\thr[][][]}]\) is
\begin{equation}
    \label{eqn:overfulfillment_thr}
    \overindicator[\user][{\thr[][][]}] = 
        \begin{cases}
           \overrate, & \!\!\!\! \text{if } \thr[\user]\ge\thr[\user][\req][] \\
            1,\,                      & \!\!\!\! \text{Otherwise}
        \end{cases}
        .
\end{equation}

The over-fulfillment rate \(\overrate\) is a constant between \(0\) and \(1\), representing the maximum over-fulfillment value considered for network metrics. Therefore, there are three different cases to be covered. The first represents a scenario where the effective throughput requirement is not met. The \(\overindicator[\user][{\thr[][][]}]\) assumes the value one and \(\intentdrift[\user][{\thr[][][]}]\) becomes a negative value between \(-1\) and \(0\) representing the distance to fulfill the requirement. The second case occurs when the intent requirement is satisfied and below the over-fulfilled value \(\thr[\user][\req][] (1+\overindicator[\user][{\thr[][][]}])\), then \(\overindicator[\user][{\thr[][][]}]\) assumes the value \(\overrate\), and \(\intentdrift[\user][{\thr[][][]}]\) computes a positive value between \(0\) and \(1\). The last case occurs when the effective throughput exceeds the over-fulfillment value or the buffer of the \ac{UE} \(\user\) is empty (indicating that the throughput requirement could not be fulfilled since there is not enough data available in the buffer to be sent).

The intent drift for buffer latency \(\intentdrift[\user][{\lat[][][]}][]\) in a simulation step \(\step\) for \ac{UE} \(\user\) is calculated as
\begin{equation}
    \label{eqn:intent_lat_ue}
    \intentdrift[\user][{\lat[][][]}] = 
    \begin{cases}
        \frac{\lat[\user][\req][] - \lat[\user]}
        {\totallat[\user] - \lat[\user][\req][] - \overindicator[\user][{\lat[][][]}]}, & \!\!\!\! \text{if } \lat[\user]>\lat[\user][\req][] (1-\overindicator[\user][{\lat[][][]}]) \\
        1,\,                      & \!\!\!\! \text{Otherwise}
    \end{cases}
    ,
\end{equation}
where the over-fulfilled requirement indicator \(\overindicator[\user][{\lat[][][]}]\) is
\begin{equation}
    \label{eqn:overfulfillment_lat}
    \overindicator[\user][{\lat[][][]}] = 
        \begin{cases}
           \totallat[\user] - \lat[\user][\req][] (1+\overrate), & \!\!\!\! \text{if } \lat[\user]\le\lat[\user][\req][] \\
            0,\,                      & \!\!\!\! \text{Otherwise}
        \end{cases}
        .
\end{equation}

Equations~\ref{eqn:intent_thr_ue} and \ref{eqn:intent_lat_ue} are similar since they represent distances to fulfill the intent requirement or a distance between the fulfillment and over-fulfillment cases. However, the intent drift for buffer latency considers that the buffer latency needs to be smaller than the requirements rather than greater than the effective throughput.

Finally, the intent drift for packet loss rate \(\intentdrift[\user][{\pktloss[][][]}][]\) in a simulation step \(\step\) for \ac{UE} \(\user\) is calculated similarly to the intent drift for the buffer latency since it also requires to have a packet loss rate below a given requirement:
\begin{equation}
    \label{eqn:intent_pkt_loss_ue}
    \intentdrift[\user][{\pktloss[][][]}] = 
    \begin{cases}
        \frac{\pktloss[\user][\req][] - \pktloss[\user]}
        {1 - \pktloss[\user][\req][] - \overindicator[\user][{\pktloss[][][]}]}, & \!\!\!\! \text{if } \pktloss[\user]>\pktloss[\user][\req][] (1-\overindicator[\user][{\pktloss[][][]}]) \\
        1,\,                      & \!\!\!\! \text{Otherwise}
    \end{cases}
    ,
\end{equation}
where the over-fulfilled requirement indicator \(\overindicator[\user][{\pktloss[][][]}]\) is
\begin{equation}
    \label{eqn:overfulfillment_pkt_loss}
    \overindicator[\user][{\pktloss[][][]}] = 
        \begin{cases}
           1 - \pktloss[\user][\req][] (1+\overrate), & \!\!\!\! \text{if } \pktloss[\user]\le\pktloss[\user][\req][] \\
            0,\,                      & \!\!\!\! \text{Otherwise}
        \end{cases}
        .
\end{equation}

The slice intent drift for the effective throughput \(\intentdrift[\slice][{\thr[][][]}][]\), buffer latency \(\intentdrift[\slice][{\lat[][][]}][]\) and packet loss rate \(\intentdrift[\slice][{\pktloss[][][]}][]\) are defined as the average of the intent drift of the \acp{UE} assigned to the slice \(\slice\)
\begin{equation}
    \label{eqn:intent_slice}
    \intentdrift[\slice][x] = \frac{\sum_{\user=1}^{\totalues[\slice]}{\intentdrift[\user][x]}}{\totalues[\slice]}, \text{for } x={\thr[][][]},{\lat[][][]}\ \text{or}\  {\pktloss[][][]}.
\end{equation}

\subsection{Inter-slice scheduler}

\subsubsection{Observation space}
\label{subsubsec:inter_obs_space}

The assumption of dealing with different network scenarios becomes a challenge to the \ac{RL} design since the observation space needs to represent different network slice-type combinations and their intents, as depicted in Fig.~\ref{fig:network_scenarios}. The proposed \ac{MARL} agent utilizes a fully connected neural network for each \ac{RL} agent with a fixed input size, and to deal with different network scenarios, it represents each type of slice by the same set of variables. We ensure the interchangeability of the observation space by representing the different types of slices with the same number of variables. To simplify scalability and generalizability, we design the observation space with always the same number of required inputs in the \ac{RL} agent, even when changing the number of active slices or slice types in the network. We also consider the same metric position for each slice in the observation space, ensuring that each input of the \ac{RL} agent is always connected to the same metric. Therefore, the meaning of each input is kept the same even when changing the network scenario.

We define the observation space of the inter-slice scheduler as
\begin{equation}
    \label{eqn:obs_space_inter}
    \obsspace[][\interslice]=[\slicemet[1][\interslice], \slicemet[2][\interslice], \dots, \slicemet[\totalslices][\interslice]]    
\end{equation}
where each \(\slicemet[\slice][\interslice]\) represents the metrics for slice \(\slice\) as a vector with common slice metrics represented by:
\begin{equation}
\begin{split}
\slicemet[\slice][\interslice] = [ \activemetric[\slice][{\thr[][][]}][]\intentdrift[\slice][{\thr[][][]}][], \activemetric[\slice][{\lat[][][]}][]\intentdrift[\slice][{\lat[][][]}][], \activemetric[\slice][{\pktloss[][][]}][]\intentdrift[\slice][{\pktloss[][][]}][], \activemetric[\slice][{\thr[][][]}][], \activemetric[\slice][{\lat[][][]}][], \activemetric[\slice][{\pktloss[][][]}][],\\ \quad \slicepriority, \frac{\thr[\slice][\req][]}{\thr[\max][\req][]}, \frac{\totalues[\slice]}{\totalues[\max][]}, \frac{\se[\slice][][][]}{\se[\max][][][]} ],
\end{split}
\end{equation}
with active intent indications \(\activemetric[\slice][{\thr[][][]},\ {\lat[][][]}\ \text{or}\ {\pktloss[][][]}][]\) representing a binary value which indicates if the intent requirement is active for slice \(\slice\). For example, in case \(\activemetric[\slice][{\thr[][][]}][]=1\), \(\activemetric[\slice][{\lat[][][]}][]=0\), and \(\activemetric[\slice][{\pktloss[][][]}][]=1\), slice \(\slice\) has intents for effective throughput and packet loss rate, but not buffer latency.

The number of variables in the observation space depends on the maximum number of slices \(\totalslices\) allowed in the system. We define a fixed maximum number of slices and, hence, a fixed number of entries in the observation space. So, it is possible to handle a variable number of active slices from \(2\) to \(\totalslices\).  We fill the vector \(\slicemet[\slice][\interslice]\) with zero values every time a slice \(s\) is not active. The intent drift values \(\intentdrift[\slice][]\) give information about which slice intents are unfulfilled, fulfilled, and over-fulfilled to the \ac{RL} agent, enabling a better distribution of the \acp{RB}. The active intent indicator \(\activemetric[\slice]\) shows which intents are enabled for each slice if the slice does not consider all of them. The \(\intentdrift[\slice][]\) is a normalized value between \(-1\) and \(1\) independent of the magnitude values of the effective throughput, buffer latency, and packet loss rate, but we still have to provide these magnitude indications so the \ac{RL} agent can differentiate slice types. Therefore, we include the normalized effective throughput requirement, the number of active \acp{UE}, and the average spectral efficiency value in the observation space.

Considering a fully connected neural network with multilayer perceptron used in the \ac{RL} algorithms, each entry in the observation space has a group of parameters whose values are changed during the \ac{RL} training according to the location of the entries in the observation space~\cite{charu2018neural}. Therefore, if an \ac{RL} agent is trained with an observation space \(\obsspace[][\interslice]=[\slicemet[1][\interslice], \slicemet[2][\interslice], \slicemet[3][\interslice]]\) with \(3\) slices, it may not be able to handle an observation space \(\obsspace[][\interslice]=[\slicemet[3][\interslice], \slicemet[2][\interslice], \slicemet[1][\interslice]]\) during the test phase even if the same group of slice information is fed to the neural network due to changes in the location of the entries, therefore representing a different state. 

When dealing with multiple network scenarios in which the maximum number of slices and entries in the neural network is fixed, the entry of each slice \(\slicemet[\slice][\interslice]\) in the observation space \(\obsspace[][\interslice]\) can be associated with different types of slices. Consequently, the \ac{RL} agent needs to be trained not only in the group of slice types but also in a different combination of the same group to increase the chances of performing well during the test phase. Taking into account the maximum number \(\totalslicetypes\) of types of slices and the maximum number of slices \(\totalslices\) in the network, the total number of combinations is \((\totalslicetypes+1)^{\totalslices}\). We propose ordering the slice entries \(\slicemet[\slice][\interslice]\) in the observation space \(\obsspace[][\interslice]\) according to the requested throughput. Therefore, the total number of combinations becomes \(\binom{\totalslices + \totalslicetypes - 1}{\totalslices}\), representing a reduction of \(100 \times \left( 1 - \frac{\binom{\totalslices + \totalslicetypes - 1}{\totalslices}}{(\totalslicetypes + 1)^{\totalslices}} \right)\). If we consider, for example, a maximum number of slices \(\totalslices=5\) and slice types \(\totalslicetypes=10\), the reduction using ordered entries is \(98.757\%\).

\subsubsection{Action space}
\label{subsubsec:inter_action_space}

The action space of the inter-slice scheduler \(\actionspace[][\interslice]\) has one output per slice
\begin{equation}
    \label{eqn:action_space_inter}
    \actionspace[][\interslice] = [\action[1][\interslice], \action[2][\interslice], \dots, \action[\totalslices][\interslice]],
\end{equation}
where \(\action[\slice][\interslice]\) represents an action factor for slice \(\slice\) with value in a range \([-1, 1]\) to match the output of the Gaussian distribution for continuous actions used~\cite{haarnoja2018soft}.

The proposed agent uses a mask for invalid actions~\cite{huang2020closer} to avoid selecting invalid actions since the number of active slices in a given step \(\step\) varies over time. Using the \ac{PPO} \ac{RL} method with a continuous action space, the output of the policy network is a probability distribution over the values of the action factor~\cite{schulman2017proximal}. Taking into account the maximum number of slices \(\totalslices\) in the system, there are \(\totalslices\) mean and standard deviation values. Every time a slice \(\slice\) is inactive, its associated mean and standard deviation values are set to \(-1\) and \(0\). Hence, its action factor \(\action[\slice][\interslice]\) will receive a value of \(-1\), and as a consequence, the number of allocated \acp{RB} to the slice \(\slice\) will be zero.

The number of allocated \acp{RB} is
\begin{equation}
    \label{eqn:inter_slice_action_allocated_rbs}
    \rbgvec = \roundsum\left(\frac{(\actionspace[][\interslice]+1)\rbgsavailable}{\sum_{\slice=1}^{\totalslices}{\action[\slice][\interslice]+1}}\right),
\end{equation}
with \(\roundsum\) representing a function that rounds fraction numbers to integers and checks if all \acp{RBG} were allocated. If the summation of \acp{RBG} for all slices is larger/smaller than the number of available \acp{RBG} \(\rbgsavailable\), it adds/removes one \ac{RBG} for each slice starting from the slice with the highest number of assigned \acp{RBG} to the slice with the smallest number until the number of allocated \acp{RBG} is equal to \(\rbgsavailable\). In case \(\sum_{\slice=1}^{\totalslices}{\action[\slice][\interslice]+1} = 0\), the available \acp{RBG} \(\rbgsavailable\) are equally distributed among the active slices.

\subsubsection{Reward}
\label{subsubsec:inter_slice_reward}

The main objective of the \ac{RL} agent to the inter-slice scheduler is to avoid/minimize slice intent violations by fulfilling the slice requirements defined in Equation~\ref{eqn:network_reqs}. Keeping the intent drift values \(\intentdrift[\slice][{\thr[][][]}]\), \(\intentdrift[\slice][{\lat[][][]}]\) and \(\intentdrift[\slice][{\pktloss[][][]}]\) between \(0\) and \(1\). The inter-slice reward \(\reward[][\interslice]\) function is
\begin{equation}
    \label{eqn:inter_reward}
    \reward[][\interslice] = 
    \begin{cases}
        \frac{\sum_{s\in\totalslices[\activemet]}{\reward[\slice][\intraslice]}}{\sum_{s\in\totalslices[\activemet]}{1}}, &\text{if } \checkviolations[\activemet][][][] = 0 \\
        \frac{\sum_{s\in\totalslices[\highpriorityunf]}{\reward[\slice][\intraslice]}}{\sum_{s\in\totalslices[\highpriorityunf]}{1}}-1, &\text{if } \checkviolations[\highpriority][][][] < 0 \\
        \frac{\sum_{s\in\totalslices[\activemetunf]}{\reward[\slice][\intraslice]}}{\sum_{s\in\totalslices[\activemetunf]}{1}}, &\text{Otherwise}
    \end{cases},
\end{equation}
where \(\checkviolations[\slicegroup][][][] = \sum_{s\in\totalslices[\slicegroup]}{\min(\min(\intentdrift[\slice][{\thr[][][]}][], \intentdrift[\slice][{\lat[][][]}][], \intentdrift[\slice][{\pktloss[][][]}][]), 0)}\) with \(\slicegroup\) representing the active \(\activemet\) or high-priority \(\highpriority\) slice group. The \(\totalslices[\activemetunf]\) and \(\totalslices[\highpriorityunf]\) represent the active and high-priority slices with unfulfilled intents. The intra-slice scheduler reward \(\reward[\slice][\intraslice]\) represents the intent-drift calculation for each evaluated slice \(\slice\) and this will be further explained in the upcoming subsection~\ref{subsubsec:intra_reward}.

When all network slice intents are met, the reward function considers the average of all active slices, resulting in a positive value between \(0\) and \(1\). Suppose that there are one or more high-priority slices with unfulfilled intents. In that case, the reward assumes the average reward value of the unfulfilled high-priority slices minus one, resulting in a negative value between \(-1\) and \(-2\). If there are no high-priority slice violations, the reward accounts for the average reward among the slices with unfulfilled intents (it does not include high-priority slice intents), obtaining a value between \(0\) and \(-1\). Every time a high-priority slice intent is not fulfilled, the proposed reward calculation accounts for only the high-priority intent values. Therefore, the proposed agent learns to fulfill the high-priority slices first and only after trying to reduce the distance to meet the requirements of the regular slices.

\subsection{Intra-slice scheduler}
\subsubsection{Observation space}

The observation space to the intra-slice scheduler is
\begin{equation}
    \label{eqn:obs_space_intra}
    \begin{split}
    \obsspace[\slice][\intraslice]= [\activemetric[\slice][{\thr[][][]}][]\intentdrift[\slice][{\thr[][][]}][], \activemetric[\slice][{\lat[][][]}][]\intentdrift[\slice][{\lat[][][]}][], \activemetric[\slice][{\pktloss[][][]}][]\intentdrift[\slice][{\pktloss[][][]}][], \activemetric[\slice][{\thr[][][]}][], \activemetric[\slice][{\lat[][][]}][], \activemetric[\slice][{\pktloss[][][]}][],\\ \quad \frac{\rbgsallocated[\slice]}{\rbgsavailable},\frac{\thr[\slice][\req][]}{\thr[\max][\req][]}, \frac{\totalues[\slice]}{\totalues[\max][]}, \vecbufferocc[\slice], \frac{\vecse[\slice][][][]}{\se[\max][][][]} ],
    \end{split}
\end{equation}
where \(\vecbufferocc[\slice]=[\bufferocc[1], \bufferocc[2], \dots, \bufferocc[{\totalues[\slice]}]]\) and \(\vecse[\slice][][][]=[ \se[1][][][], \se[2][][][], \dots, \se[{\totalues[\slice]}][][][]]\). The observation space includes the inter-slice scheduler decision on the number of allocated \acp{RB} to the slice \(\rbgsallocated[\slice]\).  The intra-slice scheduler's performance depends on the inter-slice scheduler's decisions since the number of \acp{RB} to distribute among the slice's \acp{UE} is limited by the inter-slice scheduler. Therefore, using the inter-slice scheduler decisions in the intra-slice observation space is important so the intra-slice scheduler can compute the best action given the \acp{RB} constraints.

\subsubsection{Action space}

The action space of the intra-slice scheduler \(\actionspace[\slice][\intraslice]\) for slice \(\slice\) has a unique output
\begin{equation}
    \label{eqn:action_space_intra}
    \actionspace[\slice][\intraslice] = [\action[\slice][\intraslice]],
\end{equation}
where \(\action[\slice][\intraslice]\) represents an action factor for slice \(\slice\) with an integer value from \(0\) to \(2\), which is mapped for round-robin, proportional-fair or maximum throughput algorithms. Therefore, the agent proposed for the intra-slice scheduler is responsible for selecting a scheduler algorithm to allocate the \acp{RB} assigned by the inter-slice scheduler to their \acp{UE}.

\subsubsection{Reward}
\label{subsubsec:intra_reward}

The reward calculation to the intra-slice scheduler from slice \(\slice\) is
\begin{equation}
    \label{eqn:intra_reward}
    \reward[\slice][\intraslice] = 
        \min(\intentdrift[\slice][{\thr[][][]}], \intentdrift[\slice][{\lat[][][]}], \intentdrift[\slice][{\pktloss[][][]}]).
\end{equation}
The proposed intra-slice scheduler minimizes the distance to fulfill its intents in each step \(\step\) when one of the slice intents is not fulfilled. If all the slice intents are fulfilled, the \ac{RL} maximizes the intent-drift values. The intra-slice scheduler only has information about the associated target slice \(\slice\). Therefore, it tries to maximize its reward value independently of other slice statuses.

\subsection{Baselines}

We adapted two \ac{RRS} baselines for \ac{RAN} slicing using \ac{RL} from~\cite{polese2021colo,nahum2023intent}. It is important to emphasize that the related works contain different simulated/emulated scenario assumptions, and therefore, they cannot be directly applied in the different network scenarios as proposed in this work. Therefore, we implemented adapted reward functions from these \ac{RRS} baselines~\cite{polese2021colo,mei2021intelligent} for comparison with our proposed solution. Moreover, we also implement an adaptation of the \ac{PF} and \ac{RR} algorithms~\cite{capozzi2012downlink} using multi-agent for \ac{RAN} slicing that considers each slice as a \ac{UE}.

\subsubsection{Multi-agent round-robin} Allocates the same number of \acp{RBG} to all the slices in the inter-slice scheduler. Each intra-slice scheduler equally distributes their available \acp{RBG} among the slice \acp{UE}.

\subsubsection{Multi-agent proportional-fair} It balances maximizing the total network and provides all slices with minimal service. In the inter-slice scheduler, the \ac{PF} action is
\begin{equation}
    \label{eqn:mapf_action_space_inter}
    \actionspace[\mapf][\interslice] = [\action[1][\interslice], \action[2][\interslice], \dots, \action[\totalslices][\interslice]],
\end{equation}
where the action factor \(\action[\slice][\interslice]\) is
\begin{equation}
    \action[\slice][\interslice]=\frac{\bufferocc[\slice]\totalbuffer[\slice]}{\overline{\thr[\slice]}}
\end{equation}
and \(\overline{\thr[\slice]}\) represents the average effective throughput obtained by \acp{UE} in the slice \(\slice\). Finally, the action factors are mapped to the number of \acp{RBG} using Equation~\ref{eqn:inter_slice_action_allocated_rbs}. The intra-slice schedulers use the same process to allocate the \acp{RBG} to the slice \acp{UE} but consider the \ac{UE} metrics instead of the slice metrics.

\subsubsection{Intent-aware RRS} Utilizes an adaptation from~\cite{nahum2023intent} that was originally designed to deal with \ac{eMBB}, \ac{URLLC} and \ac{BE} slices with pre-specified intents. Since we consider a varying number of slices and intents in different network scenarios, we adapted the observation space, action space, and reward calculation to support until \(\totalslices\) slices in the system, utilizing intents for effective throughput \(\thr[\slice][\req][]\), buffer latency \(\lat[\slice][\req][]\) and packet loss rate \(\pktloss[\slice][\req][]\) instead of the pre-specified intents for \ac{eMBB}, \ac{URLLC} and \ac{mMTC}. The observation space is
\begin{equation}
    \label{eqn:intent_aware_obs_space_inter}
    \obsspace[\intentaware][\interslice]=[\slicemet[1][\interslice], \slicemet[2][\interslice], \dots, \slicemet[\totalslices][\interslice]],    
\end{equation}
where each \(\slicemet[\slice][\interslice]\) represents the metrics for slice \(\slice\) as a vector with common slice metrics represented by:
\begin{equation}
\begin{split}
\slicemet[\slice][\interslice] = [ \thr[\slice][\req][], \lat[\slice][\req][], \pktloss[\slice][\req][], \se[\slice][][], \thrcap[\slice][], \thr[\slice][], \\ \bufferocc[\slice], \lat[\slice], \pktloss[\slice][], \rcvthr[\slice][]].
\end{split}
\end{equation}

It utilizes the same action space as our proposed method described in Subsection~\ref{subsubsec:inter_action_space}. The reward function of the inter-slice scheduler is
\begin{equation}
    \label{eqn:intent_aware_inter_reward}
    \begin{split}
    \reward[\intentaware][\interslice] = 
        \frac{\sum_{s\in\totalslices[\activemet]}\sum_{x\in [{\thr[][][]},{\lat[][][]},{\pktloss[][][]}]}\rwdweigth[\slice]\min(\intentdrift[\slice][x], 0)}{\sum_{s\in\totalslices[\activemet]}\rwdweigth[\slice]},
    \end{split}
\end{equation}
with \(\rwdweigth[\slice]\) being a weight that defines the importance of intents from slice \(\slice\) concerning other slices. We define \(\rwdweigth[\slice]=2\) for high-priority slices and \(\rwdweigth[\slice]=1\) for regular slices. These weights were manually assigned in~\cite{nahum2023intent}, but it is unclear how to define them when considering more than one network scenario. We define high-priority slice intent weights as double the regular slice intent values.

\subsubsection{Sched-slicing RRS} Utilizes the adaptation from the original method~\cite{polese2021colo} presented in~\cite{nahum2023intent},
utilizing a \ac{PPO} \ac{RL} agent to perform inter-slice scheduling and \ac{RR}
algorithm for intra-slice scheduling. This method was designed to deal with \ac{eMBB} and \ac{URLLC} slices through the minimization/maximization of network metrics. Since we consider varying slices and intents, we classify each slice as \ac{eMBB} or \ac{URLLC} to apply the specified method. Therefore, we consider slices with a buffer latency requirement smaller than \SI{20}{ms} as \ac{URLLC} and slices with throughput requirements bigger than \SI{20}{Mbps} as \ac{eMBB}. All slices that meet both conditions are classified as \ac{eMBB} and \ac{URLLC}.

We utilize the same observation (Subsection~\ref{subsubsec:inter_obs_space} and action space (Subsection~\ref{subsubsec:inter_action_space}) as our proposed \ac{RRS} but using the reward function
\begin{equation}
    \reward[\schedslicing][\interslice] = \sum_{\slice\in\totalslices[\embb]}(\thrcap[\slice]) - \sum_{\slice\in\totalslices[\urllc]}(\bufferocc[\slice] \totalbuffer[\slice] \pktsize[\slice]),
\end{equation}
that maximizes the served throughput \(\thrcap[\slice]\) for \ac{eMBB} slices and minimizes the buffer occupancy \(\bufferocc[\slice]\) for \ac{URLLC}.

\section{Simulation results and analysis}
\label{sec:results}

\begin{table*}[ht]
\centering
\caption{Intents and simulation parameter characteristics for each slice type. The values were adapted from the indicated references.}
\label{tab:slice_intents}
\resizebox{\textwidth}{!}{%
\begin{tabular}{ccccccccccccc}
\hline
\multirow{2}{*}{\textbf{Slice type}} &
  \multirow{2}{*}{\textbf{High-priority}} &
  \multicolumn{3}{c|}{\textbf{Intents}} &
  \multicolumn{8}{c}{\textbf{Simulation parameters}} \\ \cline{3-13} 
 &
   &
  \textbf{Served throughput \(\thr[\slice][\req][]\)} &
  \textbf{Latency \(\lat[\slice][\req][]\)} &
  \multicolumn{1}{c|}{\textbf{Reliability \(\pktloss[\slice][\req][]\)}} &
  \textbf{UE's buffer size} &
  \textbf{UE's max buffer latency} &
  \textbf{Packet size} &
  \textbf{Mobility} &
  \textbf{Requested Traffic \(\muslice[\slice]\)} &
  \textbf{Min. number UEs} &
  \textbf{Max. number UEs} &
  \textbf{Ref} \\ \hline
Control case 2         & Yes & -        & 50 ms  & 99.999999\% & 10240 packets   & 100 ms & 8192 bits  & 0 Km/h  & 5 Mbps   & 4 & 5 & \cite{ts-3gpp-22.261}  \\
Monitoring case 1      & No  & 10 Mbps  & -      & -           & 10240 packets   & 100 ms & 8192 bits  & 72 Km/h & 10 Mbps  & 4 & 5 & \cite{ts-3gpp-22.261}  \\
Robotic surgery case 1 & Yes & 20 Mbps  & 20 ms  & 99.9999\%   & 1024000 packets & 40 ms  & 16000 bits & 0 Km/h  & 30 Mbps  & 4 & 5 & \cite{ts-3gpp-22.261}  \\
Robotic diagnosis      & No  & 15 Mbps  & 20 ms  & 99.999\%    & 1024000 packets & 40 ms  & 640 bits   & 0 Km/h  & 15 Mbps  & 4 & 5 & \cite{ts-3gpp-22.261}  \\
Medical monitoring     & No  & 10 Mbps  & 100 ms & 99.9999\%   & 10240 packets   & 200 ms & 8000 bits  & 0 Km/h  & 10 Mbps  & 4 & 5 & \cite{ts-3gpp-22.261}  \\
UAV app case 1         & Yes & 100 Mbps & 200 ms & -           & 1024000 packets & 400 ms & 65536 bits & 30 Km/h & 100 Mbps & 2 & 4 & \cite{ts-3gpp-22.125}  \\
UAV control non-VLOS   & Yes & 20 Mbps  & 140 ms & 99.99\%     & 10240 packets   & 300 ms & 65536 bits & 30 Km/h & 20 Mbps  & 4 & 5 & \cite{ts-3gpp-22.125}  \\
VR gaming              & No  & 100 Mbps & 10 ms  & 99.99\%     & 1024000 packets & 20 ms  & 65536 bits & 0 Km/h  & 100 Mbps & 2 & 4 & \cite{ts-3gpp-22.261}  \\
Cloud gaming           & No  & 50 Mbps  & 80 ms  & -           & 10240 packets   & 160 ms & 65536 bits & 0 Km/h  & 50 Mbps  & 2 & 5 & \cite{nvidia2024systemreq}  \\
Video streaming 4K     & No  & 30 Mbps  & -      & -           & 10240 packets   & 100 ms & 65536 bits & 0 Km/h  & 30 Mbps  & 2 & 5 & \cite{netflix2024internetreq} \\ \hline
\end{tabular}%
}
\end{table*}

\begin{table}[ht]
    \centering
    \caption{PPO RL hyperparameter values.}
    \label{tab:hyperparams}
    \resizebox{0.7\columnwidth}{!}{%
        \begin{tabular}{ccc}
            \hline
            \textbf{Hyperparameter} & \textbf{Value}
            \\
            \hline
            SGD minibatch size         & \(64\)
            \\
            Learning rate           & \(3\cdot10^{-4}\)
            \\
            Batch size              & \(2048\)
            \\
            Gamma             & \(0.99\)
            \\

            Number SDG iterations         & \(10\)
            \\

            Lambda      & \(0.95\)
            \\

            Clip parameter \(\epsilon\)     & \(0.2\)
            \\

            Entropy coefficient      & \(0.01\)
            \\

            Value function loss coefficient      & \(0.5\)
            \\

            Gradient clip      & \(0.5\)
            \\
            
            Network architecture    & \([64,64]\)
            \\
            \hline
        \end{tabular}%
    }
\end{table}

We implemented the proposed \ac{MARL} agent with shared parameters using Ray Rllib~\cite{ray_rllib_ppo} and \ac{RL} baselines using the Stable Baselines3 library~\cite{stablebaselines3}. The \ac{RRS} simulation was implemented using Python~\cite{nahum2024sixgradiomgmt} and the simulation of the wireless channel using the QuaDRiGa simulator~\cite{QuadrigaRep}. Table~\ref{tab:slice_intents} shows the intents and characteristics for each slice type. Table~\ref{tab:hyperparams} shows the default hyperparameter values used for the proposed \ac{MARL} method and \ac{RL} baselines using \ac{PPO} \ac{RL} method.

\subsection{Network scenario generation}
\label{subsec:network_scenario}

We define a network scenario as a combination of a specific number of active slices and slice types. The number of active slices for a network scenario \(\totalslices[\scenario]\) is a random value between \(\totalslices[\min]=3\) and \(\totalslices=5\). The network scenario generator randomly selects the slice indexes to use. For example, given a network scenario with \(\totalslices[\scenario]=4\) active slices, the slice indexes used could be 1, 3, 4, and 5 while the slice index 2 is inactive. In addition, \(\totalslices[\scenario][\highpriority]\) represents the number of high-priority slices in the scenario. Each active slice has a unique slice type randomly selected from the options in Table~\ref{tab:slice_intents}. Each slice type has a high-priority indication and at least one associated intent for served throughput, latency, and reliability. The simulation parameters indicate the characteristics of the slice type \acp{UE}: buffer size, maximum buffer latency, message size, mobility, and requested traffic. The requested traffic for each \ac{UE} of a specific slice type is a Poisson distribution with a mean equal to \(\muslice[\slice]\). Finally, the number of \acp{UE} assigned for each slice type is randomly selected between the minimum and maximum number of \acp{UE} defined in Table~\ref{tab:slice_intents}.

We consider a single-input, single-output transmission system with a unique omnidirectional antenna to the base station and \acp{UE}. We obtain channel realizations every \(\samplingtime=\SI{1}{ms}\), which is the same value considered for the \ac{TTI}. The simulation episodes last \(\episodetime=\SI{1}{s}\). The \ac{UE} position is randomly selected within a range from \SI{35}{} to \SI{250}{\meter} from the base station, moving in a random direction with speed defined according to the \ac{UE}'s slice type, but always respecting the minimum and maximum distance from the base station. The \acp{UE} can turn their direction with a probability \(P_\mathrm{turn}=0.5\) in each \SI{0.2}{ms} or in case they reach the maximum distance from the base station to avoid off-limit movements. Table~\ref{tab:simulation_parameters} shows the simulation parameters considered in the \ac{RRS} system and the channel generation.

\begin{table}[t]
\centering
\caption{Network and channel generation parameters used in the simulation.}
\label{tab:simulation_parameters}
\resizebox{0.7\columnwidth}{!}{%
\begin{tabular}{cc}
\hline
\textbf{Parameters}    & \textbf{Range}                \\ \hline
Carrier frequency (\(\fc\)) & \SI{2.6}{GHz}                       \\
Bandwidth (\(\bandwidth\))              & \SI{100}{MHz}                       \\
Transmission power     & \SI{100}{Watts}                     \\
Window interval (\(\windowssize\))    & \(10\)                            \\
\acp{RB} available (\(\rbsavailable\))          & \(135\)                           \\
\acp{RBG} available (\(\rbgsavailable\))         & \(27\)                            \\
3GPP scenario          & 38.901 Urban Macro-cell \\
Max. \# of slices (\(\totalslices\)) & 5                             \\
Max. \# of UEs (\(\totalues\))       & 25                            \\
Over-fulfillment rate (\(\overrate\))       & 0.1                            \\ \hline
\end{tabular}%
}
\end{table}

There are \(200\) randomly generated network scenarios. The first \(10\) network scenarios contain \(100\) different channel episodes each. The other \(190\) network scenarios have one channel episode each. Therefore, the simulation contains \(10 \times 100+190=1190\) \ac{RL} episodes available for training and testing. We consider three simulation scenarios to evaluate our proposed method: training for a single network scenario, generalizing for multiple network scenarios, and transfer learning for unseen network scenarios.

\subsection{Training for a single network scenario}
\label{subsec:single_network_scenario}

The proposed \ac{RL} agent and baselines are trained and tested in the same network scenario. We used the first \(10\) network scenarios that each contains \(100\) different channels. Therefore, for each network scenario containing \(\totalep=100\) episodes, the agents train over \(\eptrain=60\) and utilize \(\epval=20\) for validation and \(\eptest=20\) for testing.  In the training phase, we utilize \(\epochs=10\) epochs, representing the number of times the \ac{RL} agent trains throughout the training dataset. Each episode contains \(\stepep=1000\) steps. Therefore, the training phase for the proposed agent and the baselines contains \(\step_\mathrm{train}=\eptrain\stepep\epochs=60\cdot1000\cdot10=600000\) steps.

In each of the ten trained episodes, the agent is validated over the \(\epval=20\) episodes to evaluate the agent's capacity to generalize to different channel episodes. Each episode differs in only the \acp{UE} channel trajectories in the same network scenario. We select the agent weights from the best validation iteration since the agent needs to provide a good generalization capacity for different channel episodes. This simulation scenario assesses the capacity of \ac{RRS} methods to be trained and tested for different network scenarios since the methods are designed for specific network scenarios in related works~\cite{nahum2023intent,polese2021colo}. Here, we consider ten different network scenarios to evaluate whether the same technique could be applied to other network scenarios without changing the employed method.

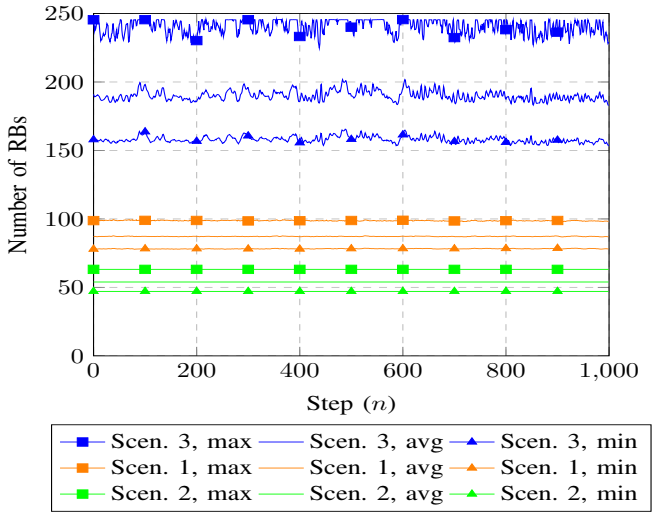
\begin{figure}[t]
    \centering
    \begin{tikzpicture}[xscale=1, yscale=0.8]
\begin{axis}[
    at={(0,0)},
    xlabel={Step (\(\step\))},
    ylabel={Number of RBs},
    xmin=0, xmax=1000,
    ymin=0, ymax=250,
    xtick distance=200,
    legend cell align=left,
    legend style={at={(0.5,-0.2)}, anchor=north, legend columns=3},
    ymajorgrids=true,
    xmajorgrids=true,
    grid style=dashed,
]

    \addplot[
        color=blue, mark=square*, mark repeat=100
    ]
    table [x=x, y=max_scenario_max, col sep=comma] {plots/data/mult_slice_seq/rbs_needed_network_scenarios.csv};
    \addlegendentry{Scen. 3, max}

    \addplot[
        color=blue
    ]
    table [x=x, y=max_scenario_avg, col sep=comma] {plots/data/mult_slice_seq/rbs_needed_network_scenarios.csv};
    \addlegendentry{Scen. 3, avg}

    \addplot[
        color=blue, mark=triangle*, mark repeat=100
    ]
    table [x=x, y=max_scenario_min, col sep=comma] {plots/data/mult_slice_seq/rbs_needed_network_scenarios.csv};
    \addlegendentry{Scen. 3, min}

    \addplot[
        color=orange, mark=square*, mark repeat=100
    ]
    table [x=x, y=median_scenario_max, col sep=comma] {plots/data/mult_slice_seq/rbs_needed_network_scenarios.csv};
    \addlegendentry{Scen. 1, max}

    \addplot[
        color=orange
    ]
    table [x=x, y=median_scenario_avg, col sep=comma] {plots/data/mult_slice_seq/rbs_needed_network_scenarios.csv};
    \addlegendentry{Scen. 1, avg}

    \addplot[
        color=orange, mark=triangle*, mark repeat=100
    ]
    table [x=x, y=median_scenario_min, col sep=comma] {plots/data/mult_slice_seq/rbs_needed_network_scenarios.csv};
    \addlegendentry{Scen. 1, min}

    \addplot[
        color=green, mark=square*, mark repeat=100
    ]
    table [x=x, y=min_scenario_max, col sep=comma] {plots/data/mult_slice_seq/rbs_needed_network_scenarios.csv};
    \addlegendentry{Scen. 2, max}

    \addplot[
        color=green
    ]
    table [x=x, y=min_scenario_avg, col sep=comma] {plots/data/mult_slice_seq/rbs_needed_network_scenarios.csv};
    \addlegendentry{Scen. 2, avg}

    \addplot[
        color=green, mark=triangle*, mark repeat=100
    ]
    table [x=x, y=min_scenario_min, col sep=comma] {plots/data/mult_slice_seq/rbs_needed_network_scenarios.csv};
    \addlegendentry{Scen. 2, min}
    
\end{axis}

\end{tikzpicture}
    \caption{Lowest, median, and highest demanding network scenarios based on the number of \acp{RB} needed to satisfy the requested traffic \(\muslice[\slice]\).}
    \label{fig:selected_scenarios_needed_rbs}
\end{figure}

From the \(10\) different network scenarios, Fig.~\ref{fig:selected_scenarios_needed_rbs} shows the lowest, median, and highest demanding network scenarios based on the number of \acp{RB} needed to satisfy the requested traffic \(\muslice[\slice]\). In each step \(\step\), it accounts for the minimum, average, and maximum spectral efficiency in the slice \acp{UE} \acp{RB} and calculates how many \acp{RB} would be needed to reach the requested traffic considering these values. Fig.~\ref{fig:selected_scenarios_needed_rbs} shows the number of required \acp{RB} to satisfy the traffic requested in each step \(\step\) from the first episode of the selected network scenario. 

The lowest demanding network scenario is the number \(2\) that needs an average number of \acp{RB} near \(53\) out of \(\rbgsavailable=135\) available. The median demanding scenario is the number \(1\), which needs an average number of \acp{RB} close to \(85\) with maximum values that get near from \(100\) \acp{RB}. The highest demanding network scenario, scenario \(3\), requires an average number of \acp{RB} close to \(190\), which surpasses the available number of \acp{RB} \(\rbgsavailable=135\), indicating that there are not sufficient resources for all slices. Therefore, the \ac{RRS} will need to prioritize the slices with higher priority. There is a low variation in the number of required \acp{RB} for scenarios \(0\) and \(1\) due to the low mobility in the selected slice types of these network scenarios. The \ac{RRS} allocates \acp{RBG}, therefore \acp{RB} are allocated in groups of \(\frac{\rbsavailable}{\rbgsavailable}=5\) \acp{RB}, as defined in Table~\ref{tab:simulation_parameters}. Table~\ref{tab:selected_scenarios} shows the slice types for each selected network scenario.

\begin{table}[t]
\centering
\caption{Slice types and number of \acp{UE} for the network scenarios \(1\), \(2\), and \(3\)}
\label{tab:selected_scenarios}
\resizebox{\columnwidth}{!}{%
\begin{tabular}{cccc}
\hline
\textbf{Slice Index} & \textbf{Scenario 1 (21 \acp{UE})}         & \textbf{Scenario 2 (13 \acp{UE})}   & \textbf{Scenario 3 (23 \acp{UE})}               \\ \hline
1 & Robotic Diagnosis (4 \acp{UE})       & Robotic surgery case 1 (5 \acp{UE}) & Monitoring case 1 (5 \acp{UE}) \\
2 & UAV control non-VLOS (5 \acp{UE}) & Medical monitoring (4 \acp{UE})       & UAV app case 1 (4 \acp{UE})   \\
3              & Cloud gaming (5 \acp{UE})       & -                     & Robotic surgery case 1 (5 \acp{UE}) \\
4              & Monitoring case 1 (5 \acp{UE}) & -                     & VR gaming (4 \acp{UE})                \\
5              & VR gaming (2 \acp{UE})          & Cloud gaming (4 \acp{UE}) & Medical monitoring (5 \acp{UE})      
\end{tabular}%
}
\end{table}

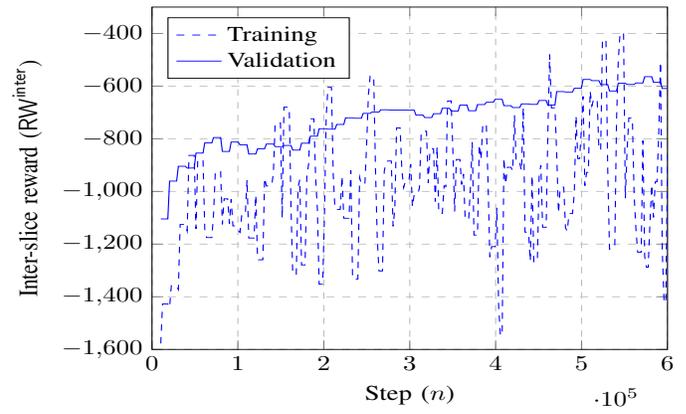
\begin{figure}[t]
    \centering
    \begin{tikzpicture}[xscale=1, yscale=0.8]
\begin{axis}[
    at={(0,0)},
    xlabel={Step (\(\step\))},
    ylabel={Inter-slice reward (\(\reward[][\interslice][]\))},
    xmin=0, xmax=600000,
    ymin=-1600, ymax=-300,
    scaled x ticks = true,
    xtick distance=100000,
    legend cell align=left,
    legend pos=north west,
    ymajorgrids=true,
    xmajorgrids=true,
    grid style=dashed,
]

    \addplot[
        color=blue, dashed
    ]
    table [x=step, y=reward, col sep=comma] {plots/data/mult_slice_seq/ray_ib_sched_default_2_train.csv};
    \addlegendentry{Training}

    \addplot[
        color=blue
    ]
    table [x=step, y=reward, col sep=comma] {plots/data/mult_slice_seq/ray_ib_sched_default_2_eval.csv};
    \addlegendentry{Validation}
    
\end{axis}

\end{tikzpicture}
    \caption{Inter-slice reward for training and validation during the \(\step_\mathrm{train}=600000\) training steps in the network scenario \(3\).}
    \label{fig:mult_slice_seq_train_val_reward}
\end{figure}

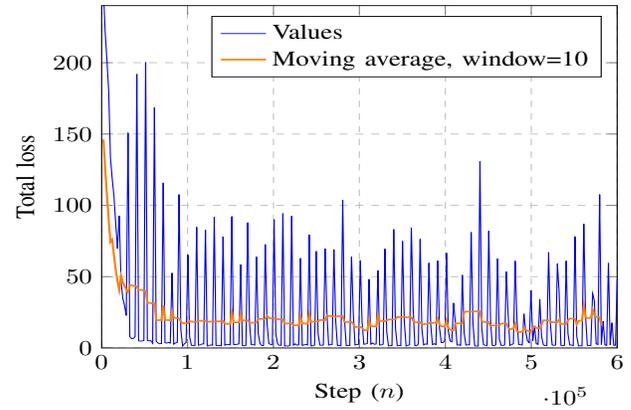
\begin{figure}[t]
    \centering
    \begin{tikzpicture}[xscale=1, yscale=0.8]
\begin{axis}[
    at={(0,0)},
    xlabel={Step (\(\step\))},
    ylabel={Total loss},
    xmin=0, xmax=600000,
    ymin=0, ymax=240,
    scaled x ticks = true,
    xtick distance=100000,
    ymajorgrids=true,
    xmajorgrids=true,
    grid style=dashed,
    legend cell align=left,
    legend pos=north east,
]

    \addplot[
        color=blue
    ]
    table [x=step, y=value, col sep=comma] {plots/data/mult_slice_seq/ray_ib_sched_default_2_loss.csv};
    \addlegendentry{Values}

    \addplot[
        color=orange, thick
    ]
    table [x=step, y=value, col sep=comma] {plots/data/mult_slice_seq/ray_ib_sched_default_2_ma_loss.csv};
    \addlegendentry{Moving average, window=10}
    
\end{axis}

\end{tikzpicture}
    \caption{Inter-slice \ac{RRS} total loss during the \(\step_\mathrm{train}=600000\) training steps in the network scenario \(3\).}
    \label{fig:mult_slice_seq_total_loss}
\end{figure}

Fig~\ref{fig:mult_slice_seq_train_val_reward} shows the inter-slice reward during training and validation to the highest demanding network scenario \(3\). We consider the inter-slice reward since it contains the contributions of all slices in the network. The training accounts for the summation of inter-slice rewards \(\reward[][\interslice][]\) in each episode, while the validation accounts for the average summation of reward values \(\reward[][\interslice][]\) in the validation episodes in every \(10\) training episodes. The proposed method improves its ability to generalize to different channel episodes over time, as depicted in the validation performance. The training performance also improves over time but has a more unstable behavior concerning the evaluation value because their values are calculated in a single episode instead of an average in a group of episodes as made in the validation. Fig~\ref{fig:mult_slice_seq_total_loss} shows the total loss to the inter-slice \ac{PPO} \ac{RL} agent. Similarly to the inter-slice rewards, the total loss also improves over training steps. The total loss still has variations over time since we are training with different channel episodes, and therefore, the policy parameters are adapted for each channel episode. The important aspect is finding a balance between the various channel episodes to reach a policy that can deal with all of them.

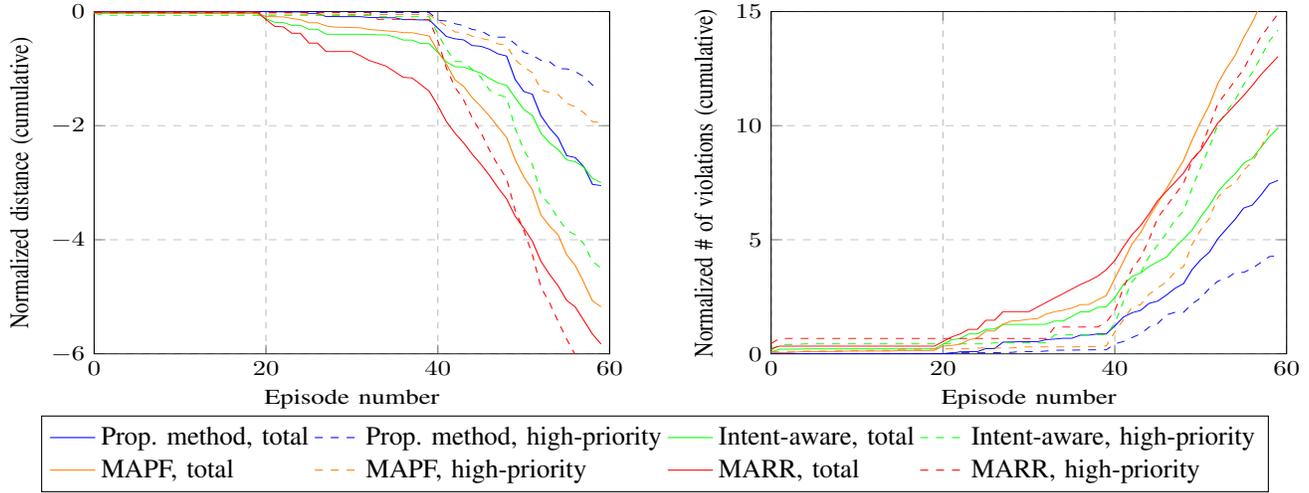
\begin{figure*}[htb]
    \centering
    \begin{tikzpicture}[xscale=1, yscale=0.8]
\begin{axis}[
    at={(0,0)},
    xlabel={Episode number},
    ylabel={Normalized distance (cumulative)},
    xmin=0, xmax=60,
    ymin=-6, ymax=0,
    xtick distance=20,
    legend cell align=left,
    legend to name={mult_slice_seq_selected_legend},
    legend style={at={(0.5,-0.1)}, anchor=north, legend columns=4},
    ymajorgrids=true,
    xmajorgrids=true,
    grid style=dashed,
]

\addplot[
    color=blue
]
table [x=x, y=ray_ib_sched_default_total, col sep=comma] {plots/data/mult_slice_seq/normalized_distance_fulfill_cumsum_selected_scenarios.csv};
\addlegendentry{Prop. method, total}

\addplot[
    color=blue, dashed
]
table [x=x, y=ray_ib_sched_default_pri, col sep=comma] {plots/data/mult_slice_seq/normalized_distance_fulfill_cumsum_selected_scenarios.csv};
\addlegendentry{Prop. method, high-priority}

\addplot[
    color=green
]
table [x=x, y=sched_twc_total, col sep=comma] {plots/data/mult_slice_seq/normalized_distance_fulfill_cumsum_selected_scenarios.csv};
\addlegendentry{Intent-aware, total}

\addplot[
    color=green, dashed
]
table [x=x, y=sched_twc_pri, col sep=comma] {plots/data/mult_slice_seq/normalized_distance_fulfill_cumsum_selected_scenarios.csv};
\addlegendentry{Intent-aware, high-priority}

\addplot[
    color=orange
]
table [x=x, y=mapf_total, col sep=comma] {plots/data/mult_slice_seq/normalized_distance_fulfill_cumsum_selected_scenarios.csv};
\addlegendentry{MAPF, total}

\addplot[
    color=orange, dashed
]
table [x=x, y=mapf_pri, col sep=comma] {plots/data/mult_slice_seq/normalized_distance_fulfill_cumsum_selected_scenarios.csv};
\addlegendentry{MAPF, high-priority}

\addplot[
    color=red
]
table [x=x, y=marr_total, col sep=comma] {plots/data/mult_slice_seq/normalized_distance_fulfill_cumsum_selected_scenarios.csv};
\addlegendentry{MARR, total}

\addplot[
    color=red, dashed
]
table [x=x, y=marr_pri, col sep=comma] {plots/data/mult_slice_seq/normalized_distance_fulfill_cumsum_selected_scenarios.csv};
\addlegendentry{MARR, high-priority}
\end{axis}

\begin{axis}[
    at={(9cm,0)},
    xlabel={Episode number},
    ylabel={Normalized \# of violations (cumulative)},
    xmin=0, xmax=60,
    ymin=0, ymax=15,
    xtick distance=20,
    ymajorgrids=true,
    xmajorgrids=true,
    grid style=dashed,
]

\addplot[
    color=blue
]
table [x=x, y=ray_ib_sched_default_total, col sep=comma] {plots/data/mult_slice_seq/normalized_violations_per_episode_cumsum_selected_scenarios.csv};
\addlegendentry{Prop. method, total}

\addplot[
    color=blue, dashed
]
table [x=x, y=ray_ib_sched_default_pri, col sep=comma] {plots/data/mult_slice_seq/normalized_violations_per_episode_cumsum_selected_scenarios.csv};
\addlegendentry{Prop. method, high-priority}

\addplot[
    color=green
]
table [x=x, y=sched_twc_total, col sep=comma] {plots/data/mult_slice_seq/normalized_violations_per_episode_cumsum_selected_scenarios.csv};
\addlegendentry{Intent-aware, total}

\addplot[
    color=green, dashed
]
table [x=x, y=sched_twc_pri, col sep=comma] {plots/data/mult_slice_seq/normalized_violations_per_episode_cumsum_selected_scenarios.csv};
\addlegendentry{Intent-aware, high-priority}

\addplot[
    color=orange
]
table [x=x, y=mapf_total, col sep=comma] {plots/data/mult_slice_seq/normalized_violations_per_episode_cumsum_selected_scenarios.csv};
\addlegendentry{MAPF, total}

\addplot[
    color=orange, dashed
]
table [x=x, y=mapf_pri, col sep=comma] {plots/data/mult_slice_seq/normalized_violations_per_episode_cumsum_selected_scenarios.csv};
\addlegendentry{MAPF, high-priority}

\addplot[
    color=red
]
table [x=x, y=marr_total, col sep=comma] {plots/data/mult_slice_seq/normalized_violations_per_episode_cumsum_selected_scenarios.csv};
\addlegendentry{MARR, total}

\addplot[
    color=red, dashed
]
table [x=x, y=marr_pri, col sep=comma] {plots/data/mult_slice_seq/normalized_violations_per_episode_cumsum_selected_scenarios.csv};
\addlegendentry{MARR, high-priority}
\legend{}
\end{axis}

\end{tikzpicture}
    \ref*{mult_slice_seq_selected_legend}
    \caption{Normalized distance to fulfill intents and number of violations to the lowest, median, and highest demanding network scenarios. The proposed method and \ac{RL} baselines train over \(\eptrain=60\) episodes and utilize \(\epval=20\) for validation and \(\eptest=20\) in each network scenario.}
    \label{fig:selected_scenarios_seq_distance_violations}
\end{figure*}

\begin{figure*}[htb]
    \centering
    \begin{tikzpicture}[xscale=1, yscale=0.8]
\begin{axis}[
    at={(0,0)},
    xlabel={Episode number},
    ylabel={Normalized distance (cumulative)},
    xmin=0, xmax=200,
    ymin=-12, ymax=0,
    xtick distance=20,
    legend cell align=left,
    legend to name={mult_slice_seq_legend},
    legend style={at={(0.5,-0.1)}, anchor=north, legend columns=4},
    ymajorgrids=true,
    xmajorgrids=true,
    grid style=dashed,
]
    
    \addplot[
        color=blue
    ]
    table [x=x, y=ray_ib_sched_default_total, col sep=comma] {plots/data/mult_slice_seq/normalized_distance_fulfill_cumsum.csv};
    \addlegendentry{Prop. method, total}

    \addplot[
        color=blue, dashed
    ]
    table [x=x, y=ray_ib_sched_default_pri, col sep=comma] {plots/data/mult_slice_seq/normalized_distance_fulfill_cumsum.csv};
    \addlegendentry{Prop. method, high-priority}
    
    \addplot[
        color=green
    ]
    table [x=x, y=sched_twc_total, col sep=comma] {plots/data/mult_slice_seq/normalized_distance_fulfill_cumsum.csv};
    \addlegendentry{Intent-aware, total}
    
    \addplot[
        color=green, dashed
    ]
    table [x=x, y=sched_twc_pri, col sep=comma] {plots/data/mult_slice_seq/normalized_distance_fulfill_cumsum.csv};
    \addlegendentry{Intent-aware, high-priority}
    
    \addplot[
        color=orange
    ]
    table [x=x, y=mapf_total, col sep=comma] {plots/data/mult_slice_seq/normalized_distance_fulfill_cumsum.csv};
    \addlegendentry{MAPF, total}
    
    \addplot[
        color=orange, dashed
    ]
    table [x=x, y=mapf_pri, col sep=comma] {plots/data/mult_slice_seq/normalized_distance_fulfill_cumsum.csv};
    \addlegendentry{MAPF, high-priority}
    
    \addplot[
        color=red
    ]
    table [x=x, y=marr_total, col sep=comma] {plots/data/mult_slice_seq/normalized_distance_fulfill_cumsum.csv};
    \addlegendentry{MARR, total}
    
    \addplot[
        color=red, dashed
    ]
    table [x=x, y=marr_pri, col sep=comma] {plots/data/mult_slice_seq/normalized_distance_fulfill_cumsum.csv};
    \addlegendentry{MARR, high-priority}
\end{axis}

\begin{axis}[
    at={(9cm,0)},
    xlabel={Episode number},
    ylabel={Normalized \# of violations (cumulative)},
    xmin=0, xmax=200,
    ymin=0, ymax=30,
    xtick distance=20,
    ymajorgrids=true,
    xmajorgrids=true,
    grid style=dashed,
]

    \addplot[
        color=blue
    ]
    table [x=x, y=ray_ib_sched_default_total, col sep=comma] {plots/data/mult_slice_seq/normalized_violations_per_episode_cumsum.csv};
    \addlegendentry{Prop. method, total}
    
    \addplot[
        color=blue, dashed
    ]
    table [x=x, y=ray_ib_sched_default_pri, col sep=comma] {plots/data/mult_slice_seq/normalized_violations_per_episode_cumsum.csv};
    \addlegendentry{Prop. method, high-priority}
    
    \addplot[
        color=green
    ]
    table [x=x, y=sched_twc_total, col sep=comma] {plots/data/mult_slice_seq/normalized_violations_per_episode_cumsum.csv};
    \addlegendentry{Intent-aware, total}
    
    \addplot[
        color=green, dashed
    ]
    table [x=x, y=sched_twc_pri, col sep=comma] {plots/data/mult_slice_seq/normalized_violations_per_episode_cumsum.csv};
    \addlegendentry{Intent-aware, high-priority}
    
    \addplot[
        color=orange
    ]
    table [x=x, y=mapf_total, col sep=comma] {plots/data/mult_slice_seq/normalized_violations_per_episode_cumsum.csv};
    \addlegendentry{MAPF, total}
    
    \addplot[
        color=orange, dashed
    ]
    table [x=x, y=mapf_pri, col sep=comma] {plots/data/mult_slice_seq/normalized_violations_per_episode_cumsum.csv};
    \addlegendentry{MAPF, high-priority}
    
    \addplot[
        color=red
    ]
    table [x=x, y=marr_total, col sep=comma] {plots/data/mult_slice_seq/normalized_violations_per_episode_cumsum.csv};
    \addlegendentry{MARR, total}
    
    \addplot[
        color=red, dashed
    ]
    table [x=x, y=marr_pri, col sep=comma] {plots/data/mult_slice_seq/normalized_violations_per_episode_cumsum.csv};
    \addlegendentry{MARR, high-priority}
    \legend{}
\end{axis}

\end{tikzpicture}
    \ref*{mult_slice_seq_legend}
    \caption{Normalized distance to fulfill intents and number of violations considering ten different network scenarios. The proposed method and \ac{RL} baselines train over \(\eptrain=60\) episodes and utilize \(\epval=20\) for validation and \(\eptest=20\) in each network scenario.}
    \label{fig:mult_slice_seq_distance_violations}
\end{figure*}
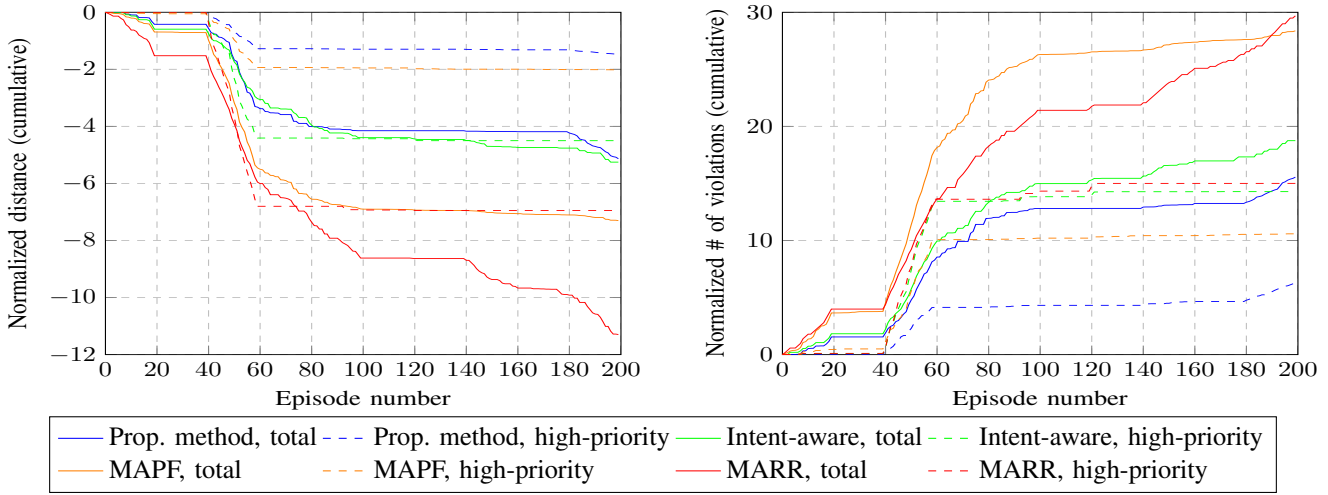

Fig.~\ref{fig:selected_scenarios_seq_distance_violations} shows the normalized distance to fulfill the slice intents of the network scenario and the normalized number of slice violations for each test episode considering the lowest, median, and highest demanding network scenarios. The number of active slices \(\totalslices[\scenario]\) normalizes the results of the scenario when considering all active slices (total) and \(\totalslices[\scenario][\highpriority]\) when considering high-priority slices. This normalization facilitates the comparison between scenarios with different numbers of active slices. Each network scenario has results for the \(\eptest=20\) episodes tested. The first \(20\) episodes represent the result for the lowest demanding scenario, the median demanding scenario from episode \(20\) to \(39\), and the highest demanding scenario from episodes \(40\) to \(59\).

The normalized distance to fulfill is the inter-slice reward (Equation~\ref{eqn:inter_reward}) but considering zero values when all the slices are fulfilled. Therefore, zero is the maximum value obtained in a simulation step \(\step\). The interpretation is how far the worst intent metric is from fulfilling its requirement. In the lowest-demand scenario, the proposed method and the baselines kept a zero distance, indicating the fulfillment of all slice intents. In the median scenario (episodes \(20\) to \(39\)), the methods start to account for values different from zero, indicating that not all the intents are fulfilled at every step of the episodes. Still, the proposed method registers the smaller distance to fulfill the high-priority and total slices. 

In the scenario with the highest demand (episodes \(40\) to \(59\)), the number of available \acp{RB} \(\rbgsavailable=135\) is insufficient to meet all the intent requirements. In this case, the \ac{RRS} methods should first satisfy the high-priority slices and then the others. The proposed method presented more robust results when considering high-priority slice protection with a smaller cumulative distance to fulfill the requirements. Due to the high priority preference, the regular slices increased their distance, as the number of \acp{RB} available is insufficient to fulfill all intents. The boolean indication of high priority incorporated in the observation and reward calculation (Subsection~\ref{subsubsec:inter_obs_space} and~\ref{subsubsec:inter_slice_reward}) of the proposed method provides better performance in protecting high-priority slices even in different network scenarios compared to the weight-based method used in the Intent-aware \ac{RRS}~\cite{nahum2023intent}. Still, the proposed method obtained the second-best performance when considering all slices with a performance close to the Intent-aware baseline.

Fig~\ref{fig:selected_scenarios_seq_distance_violations} also shows the normalized number of violations, where a slice violation occurs every time one or more slice intents are not fulfilled. Slice violation indicates a break in the \ac{SLA} while the distance to achieve intents indicates how close the unfulfilled slices are to fulfilling their requirements when there is a slice violation. The proposed method obtained the best performance for high-priority slices and total slices. The distance to fulfill intents accounts for the distance of unfulfilled slices; still, the number of fulfilled slices is higher when using the proposed method while minimizing the high-priority slice violations. This explains why it obtained the best violation results concerning all slices, although it was the second-best in the normalized distance.

Fig.~\ref{fig:mult_slice_seq_distance_violations} shows the normalized distance to fulfill the slice intents and the normalized number of slice violations, but now concerning the ten different network scenarios. Each network scenario has \(\eptest=20\) test episodes, totaling \(200\) episodes. Again, the proposed method obtained the best performance for high-priority slices in the normalized distance to fulfill and the number of violations, representing an improvement of \(40\%\) in the number of violations in relation to the baselines. In addition, it also obtained the best performance in the normalized distance and number of violations for all slices with an improvement of \(20\%\) in the number of violations.

The Sched-slicing \ac{RRS} baseline was omitted from the previous results due to its poor results in the tested network scenarios. The cumulative normalized number of violations obtained in the same simulation of Fig.~\ref{fig:mult_slice_seq_distance_violations} was \(-16\) and \(-30\) for the high-priority and total slices, which represents the highest number of violations compared to the other methods. In~\cite{nahum2023intent}, the simulation results were limited to one network scenario with one \ac{eMBB}, one \ac{URLLC}, and one \ac{mMTC} slices. The result of the Sched-slicing \ac{RRS} baseline was worse than the proposed method due to its inability to deal with the network intents since it was designed to maximize and minimize metrics and not fulfill intents. Here, we adapted the Sched-slicing \ac{RRS} baseline to deal with different network scenarios, making this approach even more difficult. When considering an intent-based network, the \ac{RRS} to be adopted must be specifically designed to deal with slice intents.

When trained for each network scenario, the proposed method performed best both in protecting high-priority and regular slices and minimizing the total number of violations in different network scenarios. It is suitable for future mobile networks because of its ability to deal with many network scenarios, simplifying the need for specific algorithms for each network scenario. In addition, the intent-based approach enables the use of the proposed method in an intent-based network to deal with high-level intents and provide the intent manager in the RAN domain a capability of fulfilling local \ac{RAN} objectives. Still, training a \ac{RL} \ac{RRS} from scratch for each network scenario can be time-consuming, and general performance could be improved by using previously learned experiences from other network scenarios. Therefore, alternatives to speed up the training of the proposed method are vital to reduce the time to deploy a new \ac{RRS} policy.

\subsection{Generalizing for multiple network scenarios}
\label{subsec:generalize_mult_scenarios}

\begin{figure}[t]
    \centering
    \begin{tikzpicture}[xscale=1, yscale=0.8]
\begin{axis}[
    at={(0,0)},
    xlabel={Step (\(\step\))},
    ylabel={Inter-slice reward (\(\reward[][\interslice][]\))},
    xmin=0, xmax=700000,
    ymin=-1900, ymax=400,
    scaled x ticks = true,
    xtick distance=100000,
    legend cell align=left,
    legend pos=north west,
    ymajorgrids=true,
    xmajorgrids=true,
    grid style=dashed,
]

    \addplot[
        color=blue, dashed
    ]
    table [x=step, y=reward, col sep=comma] {plots/data/mult_slice/ray_ib_sched_default_0_train.csv};
    \addlegendentry{Training}

    \addplot[
        color=blue
    ]
    table [x=step, y=reward, col sep=comma] {plots/data/mult_slice/ray_ib_sched_default_0_eval.csv};
    \addlegendentry{Validation}
    
\end{axis}

\end{tikzpicture}
    \caption{Inter-slice reward for training and validation during the \(\step_\mathrm{train}=900000\) training steps.}
    \label{fig:mult_slice_train_val_reward}
\end{figure}
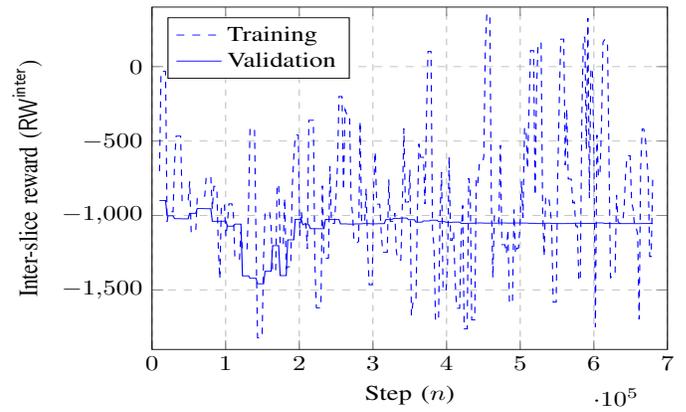

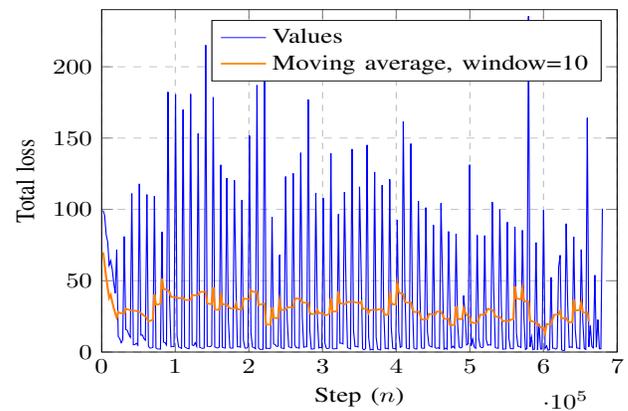
\begin{figure}[t]
    \centering
    \begin{tikzpicture}[xscale=1, yscale=0.8]
\begin{axis}[
    at={(0,0)},
    xlabel={Step (\(\step\))},
    ylabel={Total loss},
    xmin=0, xmax=700000,
    ymin=0, ymax=240,
    scaled x ticks = true,
    xtick distance=100000,
    ymajorgrids=true,
    xmajorgrids=true,
    grid style=dashed,
    legend cell align=left,
    legend pos=north east,
]

    \addplot[
        color=blue
    ]
    table [x=step, y=value, col sep=comma] {plots/data/mult_slice/ray_ib_sched_default_0_loss.csv};
    \addlegendentry{Values}

    \addplot[
        color=orange, thick
    ]
    table [x=step, y=value, col sep=comma] {plots/data/mult_slice/ray_ib_sched_default_0_ma_loss.csv};
    \addlegendentry{Moving average, window=10}
    
\end{axis}

\end{tikzpicture}
    \caption{Inter-slice \ac{RRS} total loss during the \(\step_\mathrm{train}=900000\) training steps.}
    \label{fig:mult_slice_total_loss}
\end{figure}

The proposed \ac{MARL} agent and baselines are trained and tested in different network scenarios to evaluate their generalizability.  We generate \(200\) different network scenarios where each network scenario contains unique \ac{UE} trajectories totaling \(\totalep=200\) episodes in the simulation. The \ac{RL} agents train over \(\eptrain=180\) episodes and utilize \(\epval=10\) for validation and \(\eptest=10\) for testing.  In the training phase, we utilize \(\epochs=5\) epochs. Each episode contains \(\stepep=1000\) steps. Therefore, the training phase for the proposed agent and the baselines contains \(\step_\mathrm{train}=\eptrain\stepep\epochs=180\cdot1000\cdot5=900000\) steps. 

In each of the ten trained episodes, the agent is validated over the \(\epval=10\) to evaluate the agent's capacity to generalize to different network scenarios. Therefore, each episode differs in both the \acp{UE} channel trajectories and the network scenario. The agent parameters utilized in the test phase are selected from the best validation iteration since it gives the agent the best performance to generalize to different network scenarios. This simulation scenario assesses the capacity of \ac{RRS} methods to generalize to different and unseen network scenarios without retraining for each specific network scenario. Using an agent that does not require retraining is very convenient since there is no further action to deal with new/unseen network scenarios.

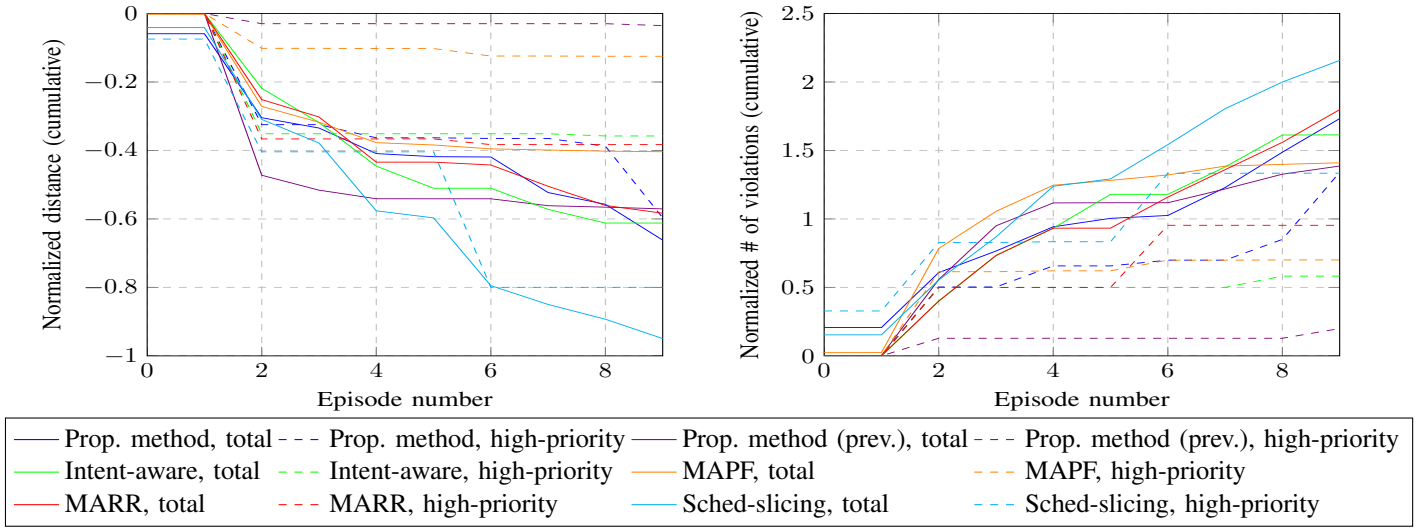
\begin{figure*}[th]
    \centering
    \begin{tikzpicture}[xscale=1, yscale=0.8]
\begin{axis}[
    at={(0,0)},
    xlabel={Episode number},
    ylabel={Normalized distance (cumulative)},
    xmin=0, xmax=9,
    ymin=-1, ymax=0,
    xtick distance=2,
    legend cell align=left,
    legend to name={mult_slice_legend},
    legend style={at={(0.5,-0.1)}, anchor=north, legend columns=4},
    ymajorgrids=true,
    xmajorgrids=true,
    grid style=dashed,
]

    \addplot[
        color=blue
    ]
    table [x=x, y=ray_ib_sched_default_total, col sep=comma] {plots/data/mult_slice/normalized_distance_fulfill_cumsum.csv};
    \addlegendentry{Prop. method, total}

    \addplot[
        color=blue, dashed
    ]
    table [x=x, y=ray_ib_sched_default_pri, col sep=comma] {plots/data/mult_slice/normalized_distance_fulfill_cumsum.csv};
    \addlegendentry{Prop. method, high-priority}

    \addplot[
        color=violet
    ]
    table [x=x, y=ray_ib_sched_default_net_sce_total, col sep=comma] {plots/data/mult_slice/normalized_distance_fulfill_cumsum.csv};
    \addlegendentry{Prop. method (prev.), total}

    \addplot[
        color=violet, dashed
    ]
    table [x=x, y=ray_ib_sched_default_net_sce_pri, col sep=comma] {plots/data/mult_slice/normalized_distance_fulfill_cumsum.csv};
    \addlegendentry{Prop. method (prev.), high-priority}




    
    \addplot[
        color=green
    ]
    table [x=x, y=sched_twc_total, col sep=comma] {plots/data/mult_slice/normalized_distance_fulfill_cumsum.csv};
    \addlegendentry{Intent-aware, total}
    
    \addplot[
        color=green, dashed
    ]
    table [x=x, y=sched_twc_pri, col sep=comma] {plots/data/mult_slice/normalized_distance_fulfill_cumsum.csv};
    \addlegendentry{Intent-aware, high-priority}
    
    \addplot[
        color=orange
    ]
    table [x=x, y=mapf_total, col sep=comma] {plots/data/mult_slice/normalized_distance_fulfill_cumsum.csv};
    \addlegendentry{MAPF, total}
    
    \addplot[
        color=orange, dashed
    ]
    table [x=x, y=mapf_pri, col sep=comma] {plots/data/mult_slice/normalized_distance_fulfill_cumsum.csv};
    \addlegendentry{MAPF, high-priority}
    
    \addplot[
        color=red
    ]
    table [x=x, y=marr_total, col sep=comma] {plots/data/mult_slice/normalized_distance_fulfill_cumsum.csv};
    \addlegendentry{MARR, total}
    
    \addplot[
        color=red, dashed
    ]
    table [x=x, y=marr_pri, col sep=comma] {plots/data/mult_slice/normalized_distance_fulfill_cumsum.csv};
    \addlegendentry{MARR, high-priority}

    \addplot[
        color=cyan
    ]
    table [x=x, y=sched_coloran_total, col sep=comma] {plots/data/mult_slice/normalized_distance_fulfill_cumsum.csv};
    \addlegendentry{Sched-slicing, total}
    
    \addplot[
        color=cyan, dashed
    ]
    table [x=x, y=sched_coloran_pri, col sep=comma] {plots/data/mult_slice/normalized_distance_fulfill_cumsum.csv};
    \addlegendentry{Sched-slicing, high-priority}
    
\end{axis}

\begin{axis}[
    at={(9cm,0)},
    xlabel={Episode number},
    ylabel={Normalized \# of violations (cumulative)},
    xmin=0, xmax=9,
    ymin=0, ymax=2.5,
    xtick distance=2,
    ymajorgrids=true,
    xmajorgrids=true,
    grid style=dashed,
]

    \addplot[
        color=blue
    ]
    table [x=x, y=ray_ib_sched_default_total, col sep=comma] {plots/data/mult_slice/normalized_violations_per_episode_cumsum.csv};
    \addlegendentry{Prop. method, total}
    
    \addplot[
        color=blue, dashed
    ]
    table [x=x, y=ray_ib_sched_default_pri, col sep=comma] {plots/data/mult_slice/normalized_violations_per_episode_cumsum.csv};
    \addlegendentry{Prop. method, high-priority}

    \addplot[
        color=violet
    ]
    table [x=x, y=ray_ib_sched_default_net_sce_total, col sep=comma] {plots/data/mult_slice/normalized_violations_per_episode_cumsum.csv};
    \addlegendentry{Prop. method (prev.), total}
    
    \addplot[
        color=violet, dashed
    ]
    table [x=x, y=ray_ib_sched_default_net_sce_pri, col sep=comma] {plots/data/mult_slice/normalized_violations_per_episode_cumsum.csv};
    \addlegendentry{Prop. method (prev.), high-priority}
    
    \addplot[
        color=green
    ]
    table [x=x, y=sched_twc_total, col sep=comma] {plots/data/mult_slice/normalized_violations_per_episode_cumsum.csv};
    \addlegendentry{Intent-aware, total}
    
    \addplot[
        color=green, dashed
    ]
    table [x=x, y=sched_twc_pri, col sep=comma] {plots/data/mult_slice/normalized_violations_per_episode_cumsum.csv};
    \addlegendentry{Intent-aware, high-priority}
    
    \addplot[
        color=orange
    ]
    table [x=x, y=mapf_total, col sep=comma] {plots/data/mult_slice/normalized_violations_per_episode_cumsum.csv};
    \addlegendentry{MAPF, total}
    
    \addplot[
        color=orange, dashed
    ]
    table [x=x, y=mapf_pri, col sep=comma] {plots/data/mult_slice/normalized_violations_per_episode_cumsum.csv};
    \addlegendentry{MAPF, high-priority}
    
    \addplot[
        color=red
    ]
    table [x=x, y=marr_total, col sep=comma] {plots/data/mult_slice/normalized_violations_per_episode_cumsum.csv};
    \addlegendentry{MARR, total}
    
    \addplot[
        color=red, dashed
    ]
    table [x=x, y=marr_pri, col sep=comma] {plots/data/mult_slice/normalized_violations_per_episode_cumsum.csv};
    \addlegendentry{MARR, high-priority}

    \addplot[
        color=cyan
    ]
    table [x=x, y=sched_coloran_total, col sep=comma] {plots/data/mult_slice/normalized_violations_per_episode_cumsum.csv};
    \addlegendentry{Sched-slicing, total}
    
    \addplot[
        color=cyan, dashed
    ]
    table [x=x, y=sched_coloran_pri, col sep=comma] {plots/data/mult_slice/normalized_violations_per_episode_cumsum.csv};
    \addlegendentry{Sched-slicing, high-priority}
    
    \legend{}
\end{axis}

\end{tikzpicture}
    \ref*{mult_slice_legend}
    \caption{Normalized distance to fulfill intents and number of violations considering \(160\) different network scenarios in the training, \(10\) in the validation, and \(10\) in the test.}
    \label{fig:mult_slice_distance_violations}
\end{figure*}

Fig~\ref{fig:mult_slice_train_val_reward} shows the inter-slice reward during training and validation. Unlike the behavior depicted in Fig~\ref{fig:mult_slice_seq_train_val_reward}, here the proposed method can hardly improve its capacity of generalizing to different network scenarios over time as shown in the validation performance, reaching its best performance in the first validation after \SI{10}{k}~trained steps. Training performance has a higher value variation that expresses instability when learning to deal with different network scenarios. Fig~\ref{fig:mult_slice_total_loss} shows the total loss to the inter-slice \ac{RL} agent. The total loss still has high values even when the number of steps increases. The training process should contain \(\step_\mathrm{train}=900000\) steps, but due to the instability in the training with high loss values, the simulation stops before the total steps.

Fig.~\ref{fig:mult_slice_distance_violations} shows the normalized distance to fulfill the slice intents and the normalized number of slice violations for \(\eptest=10\) test episodes considering ten different and unseen network scenarios. When evaluating the normalized distance, the proposed method obtained the worst performance for the high-priority slices and the second-worst performance when considering all slices. When considering the \ac{RL} baselines, they performed poorly compared to the \ac{MAPF} method. The normalized number of violations shows results similar to those of the proposed method, obtaining poor performance among the baselines.

We utilized the same ten network scenarios for testing from the results generated in the figure~\ref{fig:mult_slice_seq_distance_violations}, then it is possible to compare with the results of the proposed method trained for each specific scenario. The proposed method trained for each scenario, named "Prop. method (prev.)", shows the best performance among all the options, and not only the proposed method but all the baselines could not reach a similar performance level. These results show that the proposed method and \ac{RL} baselines cannot generalize to unseen scenarios and perform poorly compared to agents trained for each network scenario.

Since the \ac{RL}-based methods could not generalize to unseen network scenarios, we propose another experiment in which the \ac{RL} models are trained, evaluated, and tested in the same episodes in a reduced dataset. The objective is to evaluate if the \ac{RL}-based methods can overfit in the training dataset to deal with different seen network scenarios. We used \(10\) different network scenarios totaling \(\totalep=\eptrain=\epval=\eptest=10\) episodes in the simulation. The same episodes used for training are also used for validation and testing. In the training phase, we utilize \(\epochs=100\) epochs, totaling \(\step_\mathrm{train}=\eptrain\stepep\epochs=10\cdot1000\cdot100=1000000\) training steps. The objective is to overfit the proposed method and \ac{RL} baselines to evaluate whether dealing with multiple seen network scenarios is possible. The best agent weights are selected on the basis of the validation performance; in this case, the validation set is the same as the test set.

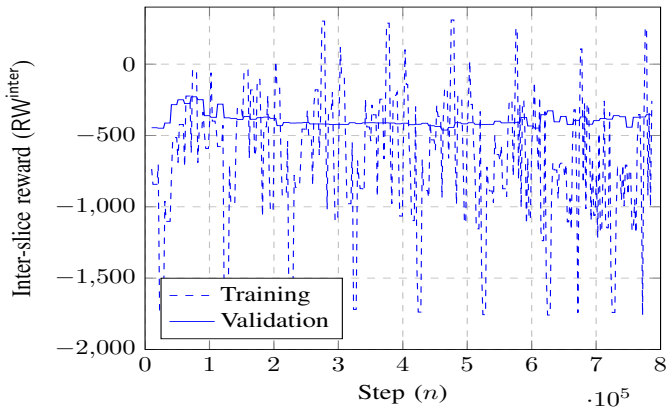
\begin{figure}[t]
    \centering
    \begin{tikzpicture}[xscale=1, yscale=0.8]
\begin{axis}[
    at={(0,0)},
    xlabel={Step (\(\step\))},
    ylabel={Inter-slice reward (\(\reward[][\interslice][]\))},
    xmin=0, xmax=800000,
    ymin=-2000, ymax=400,
    scaled x ticks = true,
    xtick distance=100000,
    legend cell align=left,
    legend pos=south west,
    ymajorgrids=true,
    xmajorgrids=true,
    grid style=dashed,
]

    \addplot[
        color=blue, dashed
    ]
    table [x=step, y=reward, col sep=comma] {plots/data/mult_slice_overfit/ray_ib_sched_default_0_train.csv};
    \addlegendentry{Training}

    \addplot[
        color=blue
    ]
    table [x=step, y=reward, col sep=comma] {plots/data/mult_slice_overfit/ray_ib_sched_default_0_eval.csv};
    \addlegendentry{Validation}
    
\end{axis}

\end{tikzpicture}
    \caption{Inter-slice reward for training and validation during the \(\step_\mathrm{train}=1000000\) training steps.}
    \label{fig:mult_slice_overfit_train_val_reward}
\end{figure}

\begin{figure}[t]
    \centering
    \begin{tikzpicture}[xscale=1, yscale=0.8]
\begin{axis}[
    at={(0,0)},
    xlabel={Step (\(\step\))},
    ylabel={Total loss},
    xmin=0, xmax=800000,
    ymin=0, ymax=150,
    scaled x ticks = true,
    xtick distance=100000,
    ymajorgrids=true,
    xmajorgrids=true,
    grid style=dashed,
    legend cell align=left,
    legend pos=north east,
]

    \addplot[
        color=blue
    ]
    table [x=step, y=value, col sep=comma] {plots/data/mult_slice_overfit/ray_ib_sched_default_0_loss.csv};
    \addlegendentry{Values}

    \addplot[
        color=orange, thick
    ]
    table [x=step, y=value, col sep=comma] {plots/data/mult_slice_overfit/ray_ib_sched_default_0_ma_loss.csv};
    \addlegendentry{Moving average, window=10}
    
\end{axis}

\end{tikzpicture}
    \caption{Inter-slice \ac{RRS} total loss during the \(\step_\mathrm{train}=1000000\) training steps.}
    \label{fig:mult_slice_overfit_total_loss}
\end{figure}
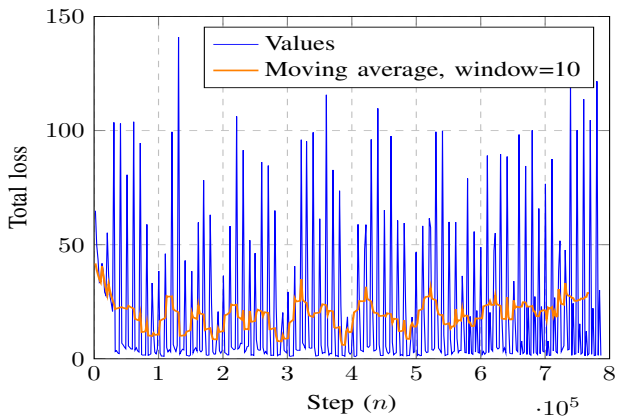

Fig~\ref{fig:mult_slice_overfit_train_val_reward} shows the inter-slice reward during training and validation. Using a smaller training set and the same set for validation and testing, the validation and training results were slightly better when compared to~\ref{fig:mult_slice_train_val_reward}. However, the proposed method cannot achieve performance similar to that demonstrated in Fig.~\ref{fig:mult_slice_seq_train_val_reward} when we train the agent for each specific network scenario. The total loss depicted in Fig.~\ref{fig:mult_slice_overfit_total_loss} obtained high values even when the training steps were increased. Again, due to the training instability, it was not able to complete the defined \(\step_\mathrm{train}=1000000\) training steps.

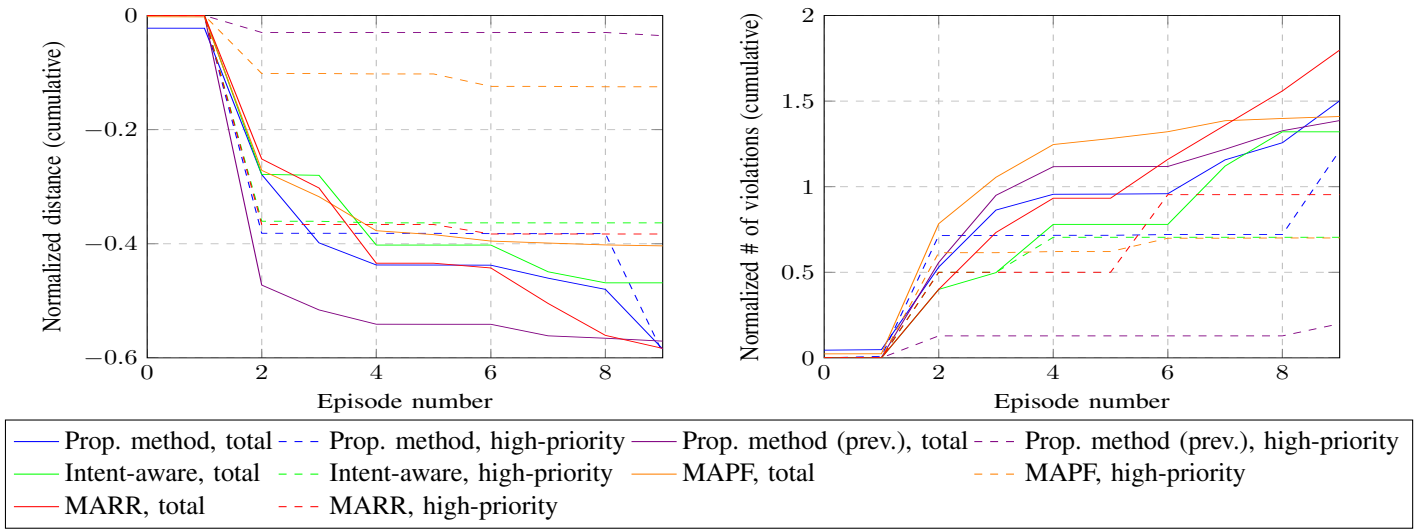
\begin{figure*}[tbh]
    \centering
    \begin{tikzpicture}[xscale=1, yscale=0.8]
\begin{axis}[
    at={(0,0)},
    xlabel={Episode number},
    ylabel={Normalized distance (cumulative)},
    xmin=0, xmax=9,
    ymin=-0.6, ymax=0,
    xtick distance=2,
    legend cell align=left,
    legend to name={mult_slice_overfit_legend},
    legend style={at={(0.5,-0.1)}, anchor=north, legend columns=4},
    ymajorgrids=true,
    xmajorgrids=true,
    grid style=dashed,
]

    \addplot[
        color=blue
    ]
    table [x=x, y=ray_ib_sched_default_total, col sep=comma] {plots/data/mult_slice_overfit/normalized_distance_fulfill_cumsum.csv};
    \addlegendentry{Prop. method, total}

    \addplot[
        color=blue, dashed
    ]
    table [x=x, y=ray_ib_sched_default_pri, col sep=comma] {plots/data/mult_slice_overfit/normalized_distance_fulfill_cumsum.csv};
    \addlegendentry{Prop. method, high-priority}

    \addplot[
        color=violet
    ]
    table [x=x, y=ray_ib_sched_default_net_sce_total, col sep=comma] {plots/data/mult_slice_overfit/normalized_distance_fulfill_cumsum.csv};
    \addlegendentry{Prop. method (prev.), total}

    \addplot[
        color=violet, dashed
    ]
    table [x=x, y=ray_ib_sched_default_net_sce_pri, col sep=comma] {plots/data/mult_slice_overfit/normalized_distance_fulfill_cumsum.csv};
    \addlegendentry{Prop. method (prev.), high-priority}




    
    \addplot[
        color=green
    ]
    table [x=x, y=sched_twc_total, col sep=comma] {plots/data/mult_slice_overfit/normalized_distance_fulfill_cumsum.csv};
    \addlegendentry{Intent-aware, total}
    
    \addplot[
        color=green, dashed
    ]
    table [x=x, y=sched_twc_pri, col sep=comma] {plots/data/mult_slice_overfit/normalized_distance_fulfill_cumsum.csv};
    \addlegendentry{Intent-aware, high-priority}
    
    \addplot[
        color=orange
    ]
    table [x=x, y=mapf_total, col sep=comma] {plots/data/mult_slice_overfit/normalized_distance_fulfill_cumsum.csv};
    \addlegendentry{MAPF, total}
    
    \addplot[
        color=orange, dashed
    ]
    table [x=x, y=mapf_pri, col sep=comma] {plots/data/mult_slice_overfit/normalized_distance_fulfill_cumsum.csv};
    \addlegendentry{MAPF, high-priority}
    
    \addplot[
        color=red
    ]
    table [x=x, y=marr_total, col sep=comma] {plots/data/mult_slice_overfit/normalized_distance_fulfill_cumsum.csv};
    \addlegendentry{MARR, total}
    
    \addplot[
        color=red, dashed
    ]
    table [x=x, y=marr_pri, col sep=comma] {plots/data/mult_slice_overfit/normalized_distance_fulfill_cumsum.csv};
    \addlegendentry{MARR, high-priority}

    
    
\end{axis}

\begin{axis}[
    at={(9cm,0)},
    xlabel={Episode number},
    ylabel={Normalized \# of violations (cumulative)},
    xmin=0, xmax=9,
    ymin=0, ymax=2,
    xtick distance=2,
    ymajorgrids=true,
    xmajorgrids=true,
    grid style=dashed,
]

    \addplot[
        color=blue
    ]
    table [x=x, y=ray_ib_sched_default_total, col sep=comma] {plots/data/mult_slice_overfit/normalized_violations_per_episode_cumsum.csv};
    \addlegendentry{Prop. method, total}
    
    \addplot[
        color=blue, dashed
    ]
    table [x=x, y=ray_ib_sched_default_pri, col sep=comma] {plots/data/mult_slice_overfit/normalized_violations_per_episode_cumsum.csv};
    \addlegendentry{Prop. method, high-priority}

    \addplot[
        color=violet
    ]
    table [x=x, y=ray_ib_sched_default_net_sce_total, col sep=comma] {plots/data/mult_slice_overfit/normalized_violations_per_episode_cumsum.csv};
    \addlegendentry{Prop. method (prev.), total}
    
    \addplot[
        color=violet, dashed
    ]
    table [x=x, y=ray_ib_sched_default_net_sce_pri, col sep=comma] {plots/data/mult_slice_overfit/normalized_violations_per_episode_cumsum.csv};
    \addlegendentry{Prop. method (prev.), high-priority}
    
    \addplot[
        color=green
    ]
    table [x=x, y=sched_twc_total, col sep=comma] {plots/data/mult_slice_overfit/normalized_violations_per_episode_cumsum.csv};
    \addlegendentry{Intent-aware, total}
    
    \addplot[
        color=green, dashed
    ]
    table [x=x, y=sched_twc_pri, col sep=comma] {plots/data/mult_slice_overfit/normalized_violations_per_episode_cumsum.csv};
    \addlegendentry{Intent-aware, high-priority}
    
    \addplot[
        color=orange
    ]
    table [x=x, y=mapf_total, col sep=comma] {plots/data/mult_slice_overfit/normalized_violations_per_episode_cumsum.csv};
    \addlegendentry{MAPF, total}
    
    \addplot[
        color=orange, dashed
    ]
    table [x=x, y=mapf_pri, col sep=comma] {plots/data/mult_slice_overfit/normalized_violations_per_episode_cumsum.csv};
    \addlegendentry{MAPF, high-priority}
    
    \addplot[
        color=red
    ]
    table [x=x, y=marr_total, col sep=comma] {plots/data/mult_slice_overfit/normalized_violations_per_episode_cumsum.csv};
    \addlegendentry{MARR, total}
    
    \addplot[
        color=red, dashed
    ]
    table [x=x, y=marr_pri, col sep=comma] {plots/data/mult_slice_overfit/normalized_violations_per_episode_cumsum.csv};
    \addlegendentry{MARR, high-priority}

    
    
    \legend{}
\end{axis}

\end{tikzpicture}
    \ref*{mult_slice_overfit_legend}
    \caption{Normalized distance to fulfill intents and number of violations considering \(10\) different network scenarios in the training and the same network scenarios for validation and test.}
    \label{fig:mult_slice_overfit_distance_violations}
\end{figure*}

Fig.~\ref{fig:mult_slice_overfit_distance_violations} shows the normalized distance to fulfill the slice intents and the normalized number of slice violations for the \(\eptest=10\) test episodes, considering that all network scenarios were seen during the training and validation phase. Even reducing the number of network scenarios from \(200\) to \(10\) and using the same dataset for training, validation, and testing, the proposed agent and \ac{RL} baselines presented poor performance compared to the proposed agent trained for each specific network scenario. The policies for each network scenario are very different, which justifies the high variation in total loss since the \ac{MARL} agent still receives large policy updates even after a considerable number of training steps. Therefore, the proposed agent cannot generalize to different network scenarios without retraining, indicating that our proposed method and baselines cannot overcome these challenges using a unique pre-trained agent to deal with all the possible network scenarios.

\subsection{Using transfer learning for unseen network scenarios}

Considering the proposed method and the baselines cannot generalize to different unseen network scenarios and do not have the capacity to handle a reduced number of trained scenarios as demonstrated in the previous Subsection~\ref{subsec:generalize_mult_scenarios}. The proposed method must be trained specifically for each network scenario. Retraining the proposed \ac{MARL} from scratch for each network scenario can take a significant amount of training steps, and the retraining frequency depends entirely on the network scenario variations faced during tests and actual deployments. Therefore, reducing the training time to achieve satisfactory performance with the proposed agent and minimizing the deployment duration in realistic environments is essential.

Due to the homogeneous observation and action space described in Subsections~\ref{subsubsec:inter_obs_space} and~\ref{subsubsec:inter_action_space}, our proposed agent can use the same neural network structures of the \ac{MARL} for different combinations of slice types and intents that characterize a network scenario. We propose using transfer learning to accelerate the training process in the requested new network scenarios and improve the performance of the proposed method. Transfer learning uses previously learned experiences while fine-tuning the \ac{RL} agent on new scenarios~\cite{zhu2023transfer}. It is usually more efficient than learning from scratch and requires less time to perform satisfactorily.

 We used the first \(10\) network scenarios containing \(100\) different channel episodes each (the same as in Section~\ref{subsec:single_network_scenario}). For each network scenario that contains \(\totalep=100\) episodes, agents train in \(\eptrain=80\) and utilize the same \(\epval=\eptest=20\) episodes for evaluation and testing. We set the same episodes for testing and evaluation to assess how many training steps agents can take to reach their best performance. In the training phase, we utilize \(\epochs=10\) epochs, totaling \(\step_\mathrm{train}=\eptrain\stepep\epochs=80\cdot1000\cdot10=800000\) trained steps. We consider the trained \ac{RL} model on Subsection~\ref{subsec:generalize_mult_scenarios} utilizing \(200\) network scenarios in the simulation as a base model for fine-tuning whose parameters are used as initial parameters for the model to be fine-tuned.

\begin{figure}[t]
    \centering
    \begin{tikzpicture}[xscale=1, yscale=0.8]
\begin{axis}[
    at={(0,0)},
    xlabel={Step (\(\step\))},
    ylabel={Inter-slice reward (\(\reward[][\interslice][]\))},
    xmin=0, xmax=800000,
    ymin=-100, ymax=300,
    scaled x ticks = true,
    xtick distance=100000,
    legend cell align=left,
    legend pos=south east,
    ymajorgrids=true,
    xmajorgrids=true,
    grid style=dashed,
]

    \addplot[
        color=blue
    ]
    table [x=step, y=reward, col sep=comma] {plots/data/finetune_mult_slice_seq/ray_ib_sched_default_0_eval.csv};
    \addlegendentry{Scratch}


    \addplot[
        color=red
    ]
    table [x=step, y=reward, col sep=comma] {plots/data/finetune_mult_slice_seq/finetune_ray_ib_sched_0_eval.csv};
    \addlegendentry{Fine-tuned}
\end{axis}

\end{tikzpicture}
    \caption{Inter-slice reward obtained in the evaluation over \(20\) episodes in the network scenario \(1\) considering a proposed agent trained from scratch with a fine-tuned agent.}
    \label{fig:finetune_mult_slice_seq_train_val_reward}
\end{figure}
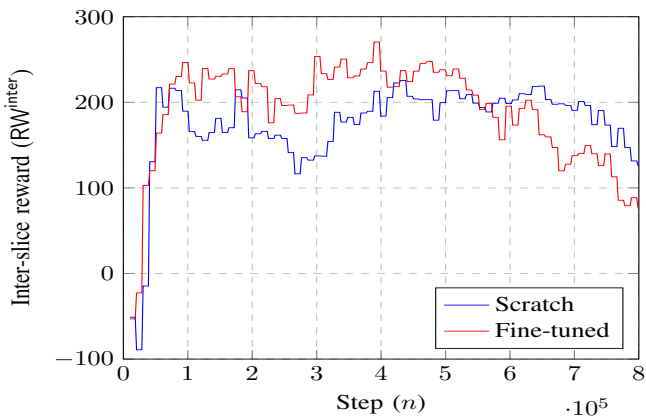

 Fig.~\ref{fig:finetune_mult_slice_seq_train_val_reward} shows the average inter-slice scheduler reward (Equation~\ref{eqn:inter_reward}) obtained in the evaluation over \(\epval=20\) episodes for the network scenario \(1\). This compares the performance of the proposed method trained from scratch with the fine-tuned agent. The proposed fine-tuned agent obtained the best performance in the evaluation considering all the trained episodes, reaching its best performance around \SI{389}{k} trained steps. There is no practical method to define how many steps the proposed agent could take to converge to its best performance, and this number of required trained steps varies according to the evaluated network scenario. To reduce the required time to deploy the method, and since there is no general number of trained steps we can ensure the convergence of the proposed method. We consider a reduction of \(8\) times in the trained steps, totaling \SI{100}{k} trained steps for analysis.
 
 When considering the best performance obtained by the proposed fine-tuned method and the proposed method trained from scratch in the first \SI{100}{k} trained steps, the proposed fine-tuned agent obtained an average reward of \(246.5\) with about \SI{92}{k} trained steps while the proposed method trained from scratch obtained an average reward value of \(217.1\) with \SI{51}{k} steps. The fine-tuned agent obtained its best average reward value (in all training episodes) of \(270.4\) with \SI{389}{k} trained steps. Therefore, the best average reward took near \(4\) times more trained steps to obtain an increase of only \(8.8\%\) in the average reward.

\begin{table*}[t]
\centering
\caption{Comparison between the proposed method trained from scratch and the fine-tuned agent in ten network scenarios over the first \(100\) and all trained episodes.}
\label{tab:fine_tune_results}
\resizebox{\textwidth}{!}{%
\begin{tabular}{|c|cccc!{\vrule width 0.3mm}cccccc|}
\hline
\multirow{3}{*}{\textbf{Scenario index}} &
  \multicolumn{4}{c!{\vrule width 0.3mm}}{\textbf{First \(100\) episodes}} &
  \multicolumn{6}{c|}{\textbf{All episodes}} \\ \cline{2-11} 
 &
  \multicolumn{2}{c|}{\textbf{Scratch}} &
  \multicolumn{2}{c!{\vrule width 0.3mm}}{\textbf{Fine-tuned}} &
  \multicolumn{3}{c|}{\textbf{Scratch}} &
  \multicolumn{3}{c|}{\textbf{Fine-tuned}} \\ \cline{2-11} 
 &
  \multicolumn{1}{c|}{\textbf{Avg. Reward}} &
  \multicolumn{1}{c|}{\textbf{Steps}} &
  \multicolumn{1}{c|}{\textbf{Avg. Reward}} &
  \textbf{Steps} &
  \multicolumn{1}{c|}{\textbf{Avg. Reward}} &
  \multicolumn{1}{c|}{\textbf{Steps}} &
  \multicolumn{1}{c|}{\textbf{Improve. (\%)}} &
  \multicolumn{1}{c|}{\textbf{Avg. Reward}} &
  \multicolumn{1}{c|}{\textbf{Steps}} &
  \textbf{Improve. (\%)} \\ \hline
1 &
  \multicolumn{1}{c|}{\cellcolor{red!25}217.1} &
  \multicolumn{1}{c|}{\cellcolor{red!25}51k} &
  \multicolumn{1}{c|}{\cellcolor{green!25}246.5} &
  \cellcolor{green!25}92k &
  \multicolumn{1}{c|}{\cellcolor{red!25}225.4} &
  \multicolumn{1}{c|}{\cellcolor{red!25}430k} &
  \multicolumn{1}{c|}{\cellcolor{red!25}3.7} &
  \multicolumn{1}{c|}{\cellcolor{green!25}270.4} &
  \multicolumn{1}{c|}{\cellcolor{green!25}389k} &
  \cellcolor{green!25}8.8 \\ \hline
2 &
  \multicolumn{1}{c|}{\cellcolor{red!25}388.8} &
  \multicolumn{1}{c|}{\cellcolor{red!25}10k} &
  \multicolumn{1}{c|}{\cellcolor{green!25}389.5} &
  \cellcolor{green!25}30k &
  \multicolumn{1}{c|}{\cellcolor{red!25}388.8} &
  \multicolumn{1}{c|}{\cellcolor{red!25}10k} &
  \multicolumn{1}{c|}{\cellcolor{red!25}0} &
  \multicolumn{1}{c|}{\cellcolor{green!25}389.9} &
  \multicolumn{1}{c|}{\cellcolor{green!25}296k} &
  \cellcolor{green!25}0.1 \\ \hline
3 &
  \multicolumn{1}{c|}{\cellcolor{red!25}-813.3} &
  \multicolumn{1}{c|}{\cellcolor{red!25}92k} &
  \multicolumn{1}{c|}{\cellcolor{green!25}-693.7} &
  \cellcolor{green!25}92k &
  \multicolumn{1}{c|}{\cellcolor{green!25}-638.1} &
  \multicolumn{1}{c|}{\cellcolor{green!25}727k} &
  \multicolumn{1}{c|}{\cellcolor{green!25}27.4} &
  \multicolumn{1}{c|}{\cellcolor{red!25}-648} &
  \multicolumn{1}{c|}{\cellcolor{red!25}409k} &
  \cellcolor{red!25}7 \\ \hline
4 &
  \multicolumn{1}{c|}{\cellcolor{red!25}-12.6} &
  \multicolumn{1}{c|}{\cellcolor{red!25}40k} &
  \multicolumn{1}{c|}{\cellcolor{green!25}-10.1} &
  \cellcolor{green!25}71k &
  \multicolumn{1}{c|}{\cellcolor{green!25}17.3} &
  \multicolumn{1}{c|}{\cellcolor{green!25}727k} &
  \multicolumn{1}{c|}{\cellcolor{green!25}173.1} &
  \multicolumn{1}{c|}{\cellcolor{red!25}6.1} &
  \multicolumn{1}{c|}{\cellcolor{red!25}747k} &
  \cellcolor{red!25}266.1 \\ \hline
5 &
  \multicolumn{1}{c|}{\cellcolor{green!25}37.2} &
  \multicolumn{1}{c|}{\cellcolor{green!25}40k} &
  \multicolumn{1}{c|}{\cellcolor{red!25}11.6} &
  \cellcolor{red!25}51k &
  \multicolumn{1}{c|}{\cellcolor{red!25}179.3} &
  \multicolumn{1}{c|}{\cellcolor{red!25}307k} &
  \multicolumn{1}{c|}{\cellcolor{red!25}79.2} &
  \multicolumn{1}{c|}{\cellcolor{green!25}187.15} &
  \multicolumn{1}{c|}{\cellcolor{green!25}358k} &
  \cellcolor{green!25}93.7 \\ \hline
6 &
  \multicolumn{1}{c|}{\cellcolor{red!25}190.7} &
  \multicolumn{1}{c|}{\cellcolor{red!25}81k} &
  \multicolumn{1}{c|}{\cellcolor{green!25}198.7} &
  \cellcolor{green!25}92k &
  \multicolumn{1}{c|}{\cellcolor{red!25}197} &
  \multicolumn{1}{c|}{\cellcolor{red!25}266k} &
  \multicolumn{1}{c|}{\cellcolor{red!25}3.1} &
  \multicolumn{1}{c|}{\cellcolor{green!25}198.7} &
  \multicolumn{1}{c|}{\cellcolor{green!25}92k} &
  \cellcolor{green!25}0 \\ \hline
7 &
  \multicolumn{1}{c|}{\cellcolor{green!25}573} &
  \multicolumn{1}{c|}{\cellcolor{green!25}30k} &
  \multicolumn{1}{c|}{\cellcolor{red!25}572.1} &
  \cellcolor{red!25}61k &
  \multicolumn{1}{c|}{\cellcolor{green!25}575.4} &
  \multicolumn{1}{c|}{\cellcolor{green!25}225k} &
  \multicolumn{1}{c|}{\cellcolor{green!25}0.4} &
  \multicolumn{1}{c|}{\cellcolor{red!25}572.6} &
  \multicolumn{1}{c|}{\cellcolor{red!25}163k} &
  \cellcolor{red!25}0 \\ \hline
8 &
  \multicolumn{1}{c|}{\cellcolor{green!25}161.1} &
  \multicolumn{1}{c|}{\cellcolor{green!25}40k} &
  \multicolumn{1}{c|}{\cellcolor{red!25}159.7} &
  \cellcolor{red!25}40k &
  \multicolumn{1}{c|}{\cellcolor{green!25}172.3} &
  \multicolumn{1}{c|}{\cellcolor{green!25}706k} &
  \multicolumn{1}{c|}{\cellcolor{green!25}6.5} &
  \multicolumn{1}{c|}{\cellcolor{red!25}161.23} &
  \multicolumn{1}{c|}{\cellcolor{red!25}194k} &
  \cellcolor{red!25}0.9 \\ \hline
9 &
  \multicolumn{1}{c|}{\cellcolor{red!25}361.9} &
  \multicolumn{1}{c|}{\cellcolor{red!25}40k} &
  \multicolumn{1}{c|}{\cellcolor{green!25}369.8} &
  \cellcolor{green!25}30k &
  \multicolumn{1}{c|}{\cellcolor{red!25}361.9} &
  \multicolumn{1}{c|}{\cellcolor{red!25}40k} &
  \multicolumn{1}{c|}{\cellcolor{red!25}0} &
  \multicolumn{1}{c|}{\cellcolor{green!25}369.8} &
  \multicolumn{1}{c|}{\cellcolor{green!25}30k} &
  \cellcolor{green!25}0 \\ \hline
10 &
  \multicolumn{1}{c|}{\cellcolor{red!25}-1097.8} &
  \multicolumn{1}{c|}{\cellcolor{red!25}92k} &
  \multicolumn{1}{c|}{\cellcolor{green!25}-1037.1} &
  \cellcolor{green!25}92k &
  \multicolumn{1}{c|}{\cellcolor{green!25}-14.2} &
  \multicolumn{1}{c|}{\cellcolor{green!25}634k} &
  \multicolumn{1}{c|}{\cellcolor{green!25}7594.6} &
  \multicolumn{1}{c|}{\cellcolor{red!25}-46.8} &
  \multicolumn{1}{c|}{\cellcolor{red!25}757k} &
  \cellcolor{red!25}2112.8 \\ \hline
\end{tabular}%
}
\end{table*}

Evaluating the results in a unique network scenario is insufficient to assess the transfer learning capacity of reducing the required steps to obtain satisfactory performance and improve overall method performance. Therefore, we summarize the results for the ten different network scenarios in Table~\ref{tab:fine_tune_results}. It presents the best average inter-slice reward value and the number of trained steps to accomplish it when considering the first \(100\) episodes and all episodes. In the first \(100\) episodes, the fine-tuned agent obtained the best performance compared to the agent trained from scratch in \(7\) network scenarios. The unique significant difference in network scenarios that the fine-tuned agent obtained a smaller average reward occurs in the network scenario \(5\). However, in the network scenarios \(7\) and \(8\), the fine-tuned and scratch agents showed a slight difference in performance.

When comparing performance in all trained episodes, the fine-tuned and scratch methods obtained the best performance in each of the \(5\) network scenarios, and the average rewards obtained had similar values, indicating that both the agent trained from scratch and the fine-tuned agent can obtain good results when trained in a large number of steps. Table~\ref{tab:fine_tune_results} also shows the percentage of improvement in the average inter-slice reward obtained when comparing the best result obtained in the first \(100\) episodes and all episodes for the fine-tuned and scratch agents. The fine-tuned agent obtained an improvement of less than \(10\%\) in \(7\) out of the \(10\) network scenarios. Indicating that in some network scenarios, training with a large number of steps may not lead to a substantial increase in the average reward obtained. However, in the network scenarios \(4\), \(5\) and \(6\), the percentage of improvement is higher than \(90\%\), obtaining \(2112\%\) in the network scenario \(10\).

The policy obtained in the generalization for multiple network scenarios (Subsection~\ref{subsec:generalize_mult_scenarios}) represents a group of common neural network parameters trained to deal with different network scenarios. Although the poor performance presented in Fig.~\ref{fig:mult_slice_distance_violations}, it is possible to interpret that the obtained policy represents an average policy to handle different network scenarios. Therefore, for most network scenarios, the parameters provided by this trained policy are far from satisfactory performance. However, it is still closer to the desired policy than the method trained from scratch. This justifies the better performance obtained in the first \(100\) trained episodes. However, it does not lead to a faster trajectory to the best parameters as presented in the comparison of steps to obtain the scratch and fine-tuned best performances in all episodes.

The proposed fine-tuned method performs best in the first \(100\) episodes and can achieve the best or near optimum performance compared to an agent trained from scratch for all episodes. Therefore, to reduce the time required to implement the \ac{RRS} for a new network scenario, the proposed method could be trained in \(100\) episodes and begin to use the agent in production. However, training in all episodes should still run in parallel, so we can substitute the production \ac{RRS} with the proposed method trained in all episodes when the training finishes to guarantee the best performance.

\section{Conclusion}
\label{sec:conclusion}

We proposed an intent-based \ac{RRS} using \ac{MARL} for inter- and intra-slice scheduling in scenarios with \ac{RAN} slicing. The \ac{RL} agent used in the inter-slice scheduler allocates the available \acp{RBG} among the slices, while the intra-slice scheduler utilizes a \ac{MARL} scheme with one \ac{RL} agent per slice, which allocates the slice \acp{RBG} to the \acp{UE}. The proposed method outperformed the baselines in protecting slices with higher priority, obtaining an improvement of \(40\%\) and, when considering all the slices, obtaining an improvement of \(20\%\) in ten different network scenarios. The results of training and testing in different network scenarios show that the proposed method and baselines cannot generalize to unseen network scenarios or even create policies to handle different trained network scenarios. We propose using transfer learning to reduce the training steps required in each network scenario. The results show that the required number of steps could be reduced by \(8\) times by using transfer learning. The proposed method first used the fine-tuned agent trained \(100\) in episodes while completing the whole training in all episodes in parallel. When the fine-tuning process is completed, we deploy the final fine-tuned agent in production to increase the method performance. Future work includes improving the model generalization for unseen network scenarios and refined transfer learning methods. Along these and other research directions, the presented evaluation methodology is useful to guide the design of \ac{RL}-based \ac{RRS} that can be deployed in practice.

\section*{Acknowledgment}
This work was partially financed by the Innovation Center, Ericsson Telecomunica\c{c}\~{o}es S.A., Brazil; Brasil 6G project (RNP/MCTI grant 01245.010604/2020-14); OpenRAN Brazil - Phase 2 project (MCTI grant Nº A01245.014203/2021-14); Universal (CNPq grant 405111/2021-5); Project Smart 5G Core And MUltiRAn Integration (SAMURAI) (MCTIC/CGI.br/FAPESP under Grant 2020/05127-2); U.S. National Science Foundation (NSF) under grants CNS-2112471, CNS-2312875 and CNS-1925601; and by OUSD(R\&E) through Army Research Laboratory Cooperative Agreement Number W911NF-19-2-0221.

\bibliographystyle{IEEEtran}
\bibliography{IEEEabrv,references}

\end{document}